\documentclass{lmcs}
\pdfoutput=1
\usepackage[utf8]{inputenc}

\usepackage{lastpage}
\lmcsdoi{21}{1}{17}
\lmcsheading{}{\pageref{LastPage}}{}{}%
{Nov.~01,~2023}{Feb.~21,~2025}{}

\pdfoutput=1

\usepackage{xspace}
\usepackage{algorithm}
\usepackage[noend]{algpseudocode}
\usepackage{amsmath}
\usepackage{amsthm}
\usepackage{amsfonts}
\usepackage{MnSymbol}
\usepackage{mathtools}
\usepackage{tikz}
\usepackage{textcomp}
\usepackage{stmaryrd}
\usepackage{varwidth}
\usepackage{graphicx}
\usepackage{mathrsfs} 

\usepackage{xcolor}
\usepackage[normalem]{ulem}
\usepackage{mdframed}
\usepackage{url}
\usepackage{wasysym}
\usepackage[outline]{contour}

\usetikzlibrary{automata, graphs,positioning,chains,arrows,snakes,decorations.pathmorphing}
\algrenewcommand\algorithmicindent{1.3em}%
\algnewcommand\algorithmicforeach{\textbf{for each}}
\algdef{S}[FOR]{ForEach}[1]{\algorithmicforeach\ #1\ \algorithmicdo}
\usepackage{todonotes}

\newcommand{\textdef}[1]{\emph{#1}}

\newmdenv[
topline=false,
bottomline=false,
rightline=false,
leftline=true,
linewidth=1.5pt,
skipabove=6mm,
skipbelow=\topsep
]{issue}

\newcommand{\cL}{\mathcal{L}}
\newcommand{\cX}{\mathcal{X}}

\newcommand{\cO}{\mathcal{O}}

\newcommand{\doc}{\operatorname{doc}}

\newcommand{\rS}{\hat{S}}

\newcommand{\cA}{\mathcal{A}}
\newcommand{\cS}{\mathcal{S}}
\newcommand{\cR}{\mathcal{R}}

\newcommand{\oalph}{\Omega}

\newcommand{\qinit}{I}

\newcommand{\ann}{\textsf{ann}}
\newcommand{\sem}[1]{{\lsem{}{#1}\rsem}}
\newcommand{\DS}{\mathcal{D}}

\newcommand{\yield}{\texttt{yield}}

\newcommand{\enumraa}{\textsc{SLPEnum}}
\newcommand{\enumedit}{\textsc{EditSLPEnum}}

\newcommand{\cE}{\mathcal{E}}

\newcommand{\hasnext}{\operatorname{hasnext}}
\newcommand{\bnodelabel}{u}

\newcommand{\leftit}{\tau_{\ell}}
\newcommand{\rightit}{\tau_{r}}

\newcommand{\ustack}{\operatorname{St}}
\newcommand{\uit}{\operatorname{\tau}}

\newcommand{\sccreate}{\textsc{create}}
\newcommand{\scnext}{\textsc{next}}
\newcommand{\scprint}{\textsc{print}}

\newcommand{\infAlph}{\Omega}
\newcommand{\D}{\mathcal{D}}
\newcommand{\cD}{\mathcal{D}}

\newcommand{\ord}{\textsf{ord}}

\newcommand{\concat}{\textsf{concat}}
\newcommand{\extract}{\textsf{extract}}
\newcommand{\delete}{\textsf{delete}}
\newcommand{\insertop}{\textsf{insert}}
\newcommand{\copyop}{\textsf{copy}}

\newcommand{\spanc}[2]{[#1,#2\rangle}

\newcommand{\dsnamebigcaps}{Enumerable Compact Set with Shifts\xspace}
\newcommand{\dsnamesbigcaps}{Enumerable Compact Sets with Shifts\xspace}

\newcommand{\dsabbr}{Shift-ECS\xspace}
\newcommand{\dsabbrs}{Shift-ECSs\xspace}

\newcommand{\shift}{\textsf{sh}}
\newcommand{\enumecs}{\textsc{ShiftECSEnum}}

\newcommand{\lch}{\ell}
\newcommand{\rch}{r}

\newcommand{\odepth}{\textsf{odepth}}
\newcommand{\add}{\textsf{add}}
\renewcommand{\prod}{\textsf{prod}}
\newcommand{\shiftop}{\textsf{shift}}
\newcommand{\union}{\textsf{union}}

\newcommand{\varop}[1]{\vdash^{#1}}
\newcommand{\varcl}[1]{\dashv^{#1}}

\newcommand{\rtname}{annotated automaton\xspace}
\newcommand{\rtnames}{annotated automata\xspace}
\newcommand{\rt}{AnnA\xspace}
\newcommand{\crt}{sAnnA\xspace}
\newcommand{\crts}{sAnnAs\xspace}
\newcommand{\rts}{AnnAs\xspace}

\newcommand{\enumE}{\mathcal{E}}
\renewcommand{\int}{\mathbb{Z}}
\newcommand{\nat}{\mathbb{N}}
\newcommand{\eps}{\varepsilon}
\renewcommand{\epsilon}{\varepsilon}

\newcommand{\br}[1]{\llbracket #1 \rrbracket}

\DeclareMathOperator{\oout}{%
	\ooalign{\scalebox{1.0}[1.0]{$o$}\cr\hidewidth\rotatebox[origin=c]{-30}{\raisebox{.4ex}{\scalebox{0.65}{$\  \boldsymbol{\smile}$}}}\hidewidth}}%

\title{Constant-delay enumeration
	for\texorpdfstring{\\}{} SLP-compressed documents}

\author[M.~Mu\~noz]{Mart\'in Mu\~noz\lmcsorcid{0009-0003-3294-6159}}	%optional
\author[C.~Riveros]{Cristian Riveros\lmcsorcid{0000-0003-0832-116X}}	%optional
\address{Pontificia Universidad Cat\'olica de Chile \& Millennium Institute for Foundational Research on Data }	%optional
\email{mmunos@uc.cl, cristian.riveros@uc.cl}  %optional

\begin{document}
	
%	\input{../response/response-letter}
%	
%	\newpage
%	
%	\setcounter{page}{1}
	
	%!TEX root = ../main.tex
% !TeX spellcheck = en_US

\begin{abstract}
We study the problem of enumerating results from a query over a compressed document. The model we use for compression are straight-line programs (SLPs), which are defined by a context-free grammar that produces a single string. For our queries, we use a model called Annotated Automata, an extension of regular automata that allows annotations on letters. This model extends the notion of regular spanners as it allows arbitrarily long outputs. Our main result is an algorithm that evaluates such a query by enumerating all results with output-linear delay after a preprocessing phase which takes linear time on the size of the SLP, and cubic time over the size of the automaton. This is an improvement over Schmid and Schweikardt's result~\cite{SchmidS21}, which, with the same preprocessing time, enumerates with a delay that is logarithmic on the size of the uncompressed document. We achieve this through a persistent data structure named Enumerable Compact Sets with Shifts which guarantees output-linear delay under certain restrictions.
These results imply constant-delay enumeration algorithms in the context of regular spanners. Further, we use an extension of annotated automata which utilizes succinctly encoded annotations to save an exponential factor from previous results that dealt with constant-delay enumeration over variable-set automata.
Lastly, we extend our results in the same fashion Schmid and Schweikardt did~\cite{SchmidS22} to allow complex document editing while maintaining the constant delay guarantee.
\end{abstract}
	
	\maketitle	
	\section{Introduction}\label{sec:introduction}
	
	%!TEX root = ../main.tex
%Strategy
%\begin{enumerate}
%	\item Relevance of Enumeration.
%	\item Document spanners and enumeration.
%	\item Extension to SLP-compressed documents. 
%	\item Drawbacks of this work.
%	\item Contributions.
%\end{enumerate}

A \emph{constant-delay enumeration algorithm} is an efficient solution to an enumeration problem: given an instance of the problem, the algorithm performs a preprocessing phase to build some indices, to then continue with an enumeration phase where it retrieves each output, one by one, taking constant time (i.e., delay) between consecutive outputs.
These algorithms provide a strong guarantee of efficiency since a user knows that, after the preprocessing phase, she will access the output as if we have already computed it. 
For these reasons, constant-delay algorithms have attracted researchers' attention, finding sophisticated solutions to several query evaluation problems. Starting with Durand and Grandjean's work~\cite{DurandG07}, researchers have found constant-delay algorithms for various classes of conjunctive queries~\cite{BaganDG07,CarmeliZBKS20}, FO queries over sparse structures~\cite{KazanaS11,SchweikardtSV18}, and MSO queries over words and trees~\cite{Bagan06,AmarilliBJM17}.

The enumeration problem over documents (i.e., strings) has been studied extensively under the framework of document spanners~\cite{FaginKRV15}. A constant-delay algorithm for evaluating deterministic regular spanners was first presented in~\cite{FlorenzanoRUVV20} and extended to non-deterministic regular spanners in~\cite{AmarilliBMN21}. After these results, people have studied the enumeration problem of document spanners in the context of ranked enumeration~\cite{DoleschalKMP22,BourhisGJR21}, nested documents~\cite{MunozR22}, or  grammars~\cite{Peterfreund21,AmarilliJMR22}. 
Recently, Schmid and Schweikardt~\cite{SchmidS21,SchmidS22} studied the evaluation problem for regular spanners over a document compressed by a Straight-Line Program (SLP). In this setting, one encodes a document through a context-free grammar that produces a single string (i.e., the document itself). This mechanism allows to highly compress documents, in some instances resulting in logarithmic space compared to the uncompressed copy.
The enumeration problem consists now of evaluating a regular spanner over an SLP-compressed document. In~\cite{SchmidS21}, the authors provided a logarithmic-delay (over the uncompressed document) algorithm for the 
%evaluation 
problem, and in~\cite{SchmidS22}, they extended this setting to edit operations over SLP documents, maintaining the delay. In particular, these works left open whether one can solve the enumeration problem of regular spanners over SLP-compressed documents with a constant-delay guarantee. 

In this paper, we extend the understanding of the evaluation problem over SLP-compressed documents in several directions. 
\begin{enumerate}
	\item We study the evaluation problem of \emph{annotated automata} (AnnA) over SLP-compressed documents. These automata are a general model for defining regular enumeration problems, which strictly generalizes the model of extended variable-set automaton used in~\cite{SchmidS21}. 
	\item We provide an output-linear delay enumeration algorithm for the problem of evaluating an unambiguous AnnA over an SLP-compressed document (Theorem~\ref{theo:main-theorem}). In particular, this result implies a constant-delay enumeration algorithm for evaluating extended variable-set automaton (formalized in Corollary~\ref{coro:eva}), giving a positive answer to the open problem left in~\cite{SchmidS21}.
	\item We show that this result extends to what we call a \emph{succinctly} annotated automaton, a generalization of AnnA whose annotations are succinctly encoded by an enumeration scheme. We develop an output-linear delay enumeration algorithm for this model (Theorem~\ref{theo-algo-succ}), and then show a constant-delay algorithm for sequential  (non-extended) variable set automata (Corollary~\ref{coro:va}), strictly generalizing the work in~\cite{SchmidS21}. 
	\item Finally, we show that one can maintain these algorithmic results when dealing with complex document editing as in~\cite{SchmidS22} (Theorem~\ref{theo:edit-main-theorem}).
\end{enumerate}

The main technical result is to show that the data structure presented in~\cite{MunozR22}, called Enumerable Compact Set (ECS), can be extended to deal with shift operators (we call it \dsabbr). This extension allows us to compactly represent the outputs and ``shift'' the results in constant time, in the sense of adding or subtracting a common value to all elements in a set. By using matrices with \dsabbr nodes, we can then perform a bottom-up evaluation of the annotated automaton over the grammar (similar to~\cite{SchmidS21}) to enumerate all outputs with output-linear delay. The combination of annotated automata and \dsabbrs considerably simplifies the algorithm's presentation, and lets us achieve a better delay bound.

\paragraph{Earlier version} This paper is an extended version of the article ``{\it Constant-delay enumeration for SLP-compressed documents}\,'' that was published in the 26th International Conference on Database Theory (ICDT 2023)~\cite{MunozR23}.
We have made several changes for this version. 
The main one is that we have included full proofs for all of our results. 
In particular, the enumeration algorithm of the Shift-ECS data structure was completely redone by introducing the use of iterators.
This change not only makes the algorithm easier to follow but also makes it more suitable for future implementation.
Most importantly, it fixes a subtle mistake that was present in the previous algorithm. 
This version also includes additional examples and explanatory figures in an effort to improve readability.

\paragraph{Organization of the paper} In Section~\ref{sec:setting} we introduce the setting and its corresponding enumeration problem. In Section~\ref{sec:ecs}, we present our data structure for storing and enumerating the outputs, and in Section~\ref{sec:ecs-op} the operations for extending it. Then, in Section~\ref{sec:evaluation} we show the evaluation algorithm. Section~\ref{sec:spanners} offers the application of the algorithmic results to document spanners, plus an extension for compressed annotation schemes, and Section~\ref{sec:edits} shows how to extend these results to deal with complex document editing. 
We finish the paper with future work in Section~\ref{sec:conclusions}.

	\section{Setting and main problem}\label{sec:setting}
	
	%!TEX root = ../main.tex

In this section, we present the setting and state the main result. 
First, we define straight-line programs, which we will use for the compressed representation of input documents. Then we introduce the definition of \rtname, an extension of regular automata to produce outputs. We use \rtnames as our computational model to represent queries over documents. 
By combining both formalisms, we state the main enumeration problem and the main technical result. %Before this, we start by providing some basic notation to work with documents. 

\paragraph{Documents} Given a finite alphabet $\Sigma$, a \emph{document} $d$ over $\Sigma$ (or just a document) is a string $d = a_1 a_2 \ldots a_n\in\Sigma^*$. Given documents $d_1$ and $d_2$, we write $d_1 \cdot d_2$ (or just $d_1d_2$) for the concatenation of $d_1$ and $d_2$. We denote by $|d| = n$ the length of the document $d = a_1 \ldots a_n$ and by $\epsilon$ the document of length $0$. 
We use $\Sigma^*$ to denote the set of all documents, and~$\Sigma^+$ for all documents with one or more symbols.
In the sequel, we will usually use $d$ for a document, and $a$ or $b$ for a symbol in $\Sigma$. 

\paragraph{SLP-compression} A \emph{context-free grammar} is a tuple $G = (N,\Sigma,R,S_0)$, where $N$ is a non-empty set of non-terminals, $\Sigma$ is finite alphabet, $S_0\in N$ is the start symbol and $R\subseteq N \times(N \cup\Sigma)^{+}$ is the set of rules. As a convention, the rule $(A,w) \in R$ will be written as $A\to w$, and we will call $\Sigma$ and $N$ the set of terminal and non-terminal symbols, respectively. A context-free grammar $S = (N,\Sigma,R,S_0)$ is a \emph{straight-line program} (SLP) if $R$ is a total function from $N$ to $(N \cup\Sigma)^{+}$ and the directed graph $(N, \{(A,B)\mid (A,w) \in R \text{ and } B\text{ appears in } w\})$ is acyclic. 
For every $A \in N$, let $R(A)$ be the unique $w \in (N \cup\Sigma)^{+}$ such that $(A,w) \in R$, and for every $a \in \Sigma$ let $R(a) = a$.
We extend $R$ to a morphism $R^*: (N \cup\Sigma)^*\to \Sigma^*$ recursively such that $R^*(d) = d$ when $d$ is a document, and $R^*(\alpha_1 \ldots \alpha_n) = R^*\big(R(\alpha_1) \cdot \ldots \cdot R(\alpha_n)\big)$, where $\alpha_i \in (N \cup \Sigma)$ for every $i \leq n$. 
By our definition of SLP, $R^*(A)$ is in $\Sigma^+$, and uniquely defined for each $A\in N$. 
Then we define the document encoded by $S$ as $\doc(S) = R^*(S_0)$.
\begin{exa}	
	Let $S = (N, \Sigma, R, S_0)$ be an SLP with $N = \{S_0, A, B\}$, $\Sigma = \{{\sf a}, {\sf b}, {\sf r}\}$, and $R = \{S_0 \to A{\sf r}BABA, A\to {\sf ba}, B \to A{\sf ra}\}$. We then have that $\doc(A) = {\sf ba}$, $\doc(B) = {\sf bara}$ and $\doc(S) = \doc(S_0) = {\sf barbarababaraba}$, namely, the string represented by $S$.%. In conclusion, $S$ is an SLP for the string ${\sf barbarababaraba}$.
\end{exa}

We define the size of an SLP $S = (N,\Sigma,R,S_0)$ as $|S|= \sum_{A \in N} |R(A)|$, namely, the sum of the lengths of the right-hand sides of all rules.
It is important to note that an SLP $S$ can encode a document $\doc(S)$ such that $|\doc(S)|$ is exponentially larger with respect to $|S|$. For this reason, SLPs are a commonly used data compression scheme~\cite{StorerS82, KiefferY00, Rytter02, ClaudeN11}, and they are often studied particularly because of their algorithmic properties; see~\cite{Lohrey12} for a survey. In this paper, we consider SLP compression to represent documents and use the formalism of annotated automata for extracting relevant information from the document.

\paragraph{Annotations} 
The outputs that we deal with in this work follow a fixed format:
for a given input alphabet $\Sigma$, a document $d = a_1\ldots a_n\in\Sigma^*$, and an output alphabet $\oalph$, an {\em annotation sequence} (or just {\em annotation}) over $d$, is a sequence $(\oout_1, i_1) \ldots (\oout_k, i_k)$ for some $k \leq n$ where $1\leq i_1  < \ldots < i_k \leq n$.
Intuitively, the sequence represents a way of annotating $d$, where each $(\oout, i)$ means that the position $i$ is annotated with the symbol~$\oout$.

These sequences are represented as strings of output-position pairs for convenience, but they could be, in principle, rearranged freely and represent the same annotation.\footnote{This distinction is what we believe separates the query model from transducers: the user is not interested in the outputs as strings, but rather as annotations.}

\paragraph{Annotated automata} 
An \emph{\rtname} (\rt for short) is a finite state automaton where we label some transitions with output symbols. Formally, it is a tuple $\cA = (Q, \Sigma, \oalph, \Delta, \qinit, F)$ where $Q$ is a state set, $\Sigma$ is a finite input alphabet, $\oalph$ is a finite output alphabet, $\qinit \subseteq Q$ and $F \subseteq Q$ are the initial and final set of states, respectively, and
\[
\Delta \ \ \subseteq \ \ \underbrace{Q \times \Sigma \times Q}_{\text{read transitions}} \ \, \cup \underbrace{Q \times (\Sigma \times \oalph) \times Q}_{\text{read-annotate transitions}}
\] 
is the transition relation, which contains {\em read transitions} of the form  $(p, a, q)\in Q\times\Sigma\times Q$, and {\em read-annotate transitions} of the form $(p, (a, \oout), q)\in Q\times(\Sigma\times\oalph)\times Q$. We define the size of $\cA$ as $|\cA| = |Q| + |\Delta|$. 

Similarly to transducers~\cite{berstel2013transductions}, a symbol $a \in \Sigma$ is an input symbol that the machine reads and $\oout \in \oalph$ is a symbol that indicates what the machine prints in an output tape.
% The key difference is that annotations are always associated to an index in the document, which is printed along them in the output string.
A run $\rho$ of $\cA$ over a document $d = a_1a_2\ldots a_n \in\Sigma^*$ is a sequence of the form:
\[
%\rho = (q_1, \sigma_1) \xrightarrow{s_1/\!\ooutscr_1} (q_2, \sigma_2) \xrightarrow{s_2/\!\ooutscr_2} \ldots  \xrightarrow{s_n/\!\ooutscr_n} (q_{n+1}, \sigma_{n+1})
\rho \ := \ q_1 \xrightarrow{b_1} q_2 \xrightarrow{b_2} \ldots \xrightarrow{b_{n+1}} q_{n+1}
\]
such that $q_1\in\qinit$ and, for each $i\in [1,n]$, it holds that either $b_i = a_i$ and $(q_i,a_i,q_{i+1})\in\Delta$, or $b_i = (a_i,\oout)$ and $(q_{i},(a_i,\oout),q_{i+1})\in\Delta$. 
We say that $\rho$ is accepting if $q_{n+1}\in F$.

We define the \emph{annotation of $\rho$} as
$
\ann(\rho) =  \ann(b_1)\cdot \ldots \cdot \ann(b_n)
$
such that $\ann(b_i) = (\oout, i)$ if $b_i = (a, \oout)$, and $\ann(b_i) = \eps$ otherwise, for each $i\in[1,n]$.
%Note that in $\nu = \oout_1 \cdots \oout_n$, $\oeps$ is a symbol that may be present in the word, whereas $\eps$ indicates an empty word. 
Given an \rtname $\cA$ and a document $d \in\Sigma^*$, we define the set $\sem{\cA}(d)$ of all outputs of $\cA$ over $d$ as:
%$$
%\sem{\cA}(w) \ = \ \{ \out(\rho) \, \mid \, \text{$\rho$ is an accepting run of $\cA$ over $w$}\}.
%$$
\[
\sem{\cA}(d) \ = \ \{ \ann(\rho) \, \mid \, \text{$\rho$ is an accepting run of $\cA$ over $d$}\}.
\] 
Note that each output in $\sem{\cA}(d)$ is an annotation sequence over $d$.

\begin{figure}[t]
	\centering
	
	\begin{tikzpicture}[scale=0.7,->,>=stealth',shorten >=1pt,auto,node distance=2cm,thick,state/.style={circle,draw}, color=black]
		\node[state,draw=none,scale=0.1] (in) at (-2,0) {};
		\node[state] (q0) at (0,0) {$q_1$};
		\node[state] (q1) at (4,0) {$q_2$};
		\node[state] (q2) at (8,0) {$q_3$};
		\node[state, accepting] (q3) at (12,0) {$q_4$};
		\draw (in) to (q0);
		\draw (q0) to[loop above] node {${\sf a}, {\sf b}, {\sf r}$} (q0);
		\draw (q0) to node [above] {$({\sf b},\circ)$} (q1);
		\draw (q1) to[loop above] node {${\sf a}$} (q1);
		\draw (q1) to node [above] {$({\sf r},\bullet)$} (q2);
		\draw (q2) to[loop above] node {${\sf a}$} (q2);
		\draw (q2) to node [above] {$({\sf b},\circ)$} (q3);
		\draw (q3) to[loop above] node {${\sf a}, {\sf b}, {\sf r}$} (q3);
	\end{tikzpicture}
	
	\caption{Example of an annotated automaton.}
	
	\label{fig-anna-ex}
\end{figure}
\begin{exa}
	Consider an \rt $\cA = (Q, \Sigma, \oalph, \Delta, \qinit, F)$ where $\qinit = \{q_1\}$ $Q = \{q_1, q_2, q_3, q_4\}$, $\Sigma = \{{\sf a}, {\sf b}, {\sf r}\}$, $\oalph = \{\circ,\bullet\}$, and $F =\{q_4\}$. We define $\Delta$ as the set of transitions that are depicted in Figure~\ref{fig-anna-ex}. 
	For the document $d = {\sf barbarababaraba}$ the set $\sem{\cA}(d)$ contains the strings $(\circ, 1)(\bullet, 3)(\circ, 4)$, $(\circ, 4)(\bullet, 6)(\circ, 8)$ and $(\circ, 10)(\bullet, 12)(\circ, 14)$. 
	In short, $\cA$ selects triples of ${\sf b}$, ${\sf r}$ and ${\sf b}$ which are separated only by ${\sf a}$, and marks them with the symbols $\circ, \bullet$ and $\circ$, respectively.
\end{exa}
Annotated automata are the natural regular counterpart of annotated grammars introduced in~\cite{AmarilliJMR22}. Moreover, it is the generalization and simplification of similar automaton formalisms introduced in the context of information extraction~\cite{FaginKRV15,Peterfreund21}, complex event processing~\cite{grez2021formal,GrezR20}, and enumeration in general~\cite{BourhisGJR21,MunozR22}. In Section~\ref{sec:spanners}, we show how we can reduce the automaton model of document spanners, called a variable-set automaton, into a (succinctly) annotated automaton, generalizing the setting in~\cite{SchmidS21}. 

Similar to other automata models, the notion of an unambiguous automaton will be crucial for us to remove duplicate runs for the same output. 
Specifically, we say that an \rt $\cA = (Q, \Sigma, \oalph, \Delta, \qinit, F)$ is 
{\it unambiguous} if for every $d \in \Sigma^*$ and every $w \in \sem{\cA}(d)$ there is exactly one accepting run $\rho$ of $\cA$ over $d$ such that $w = \ann(\rho)$. 
On the other hand, we say that $\cA$ is \emph{deterministic} if $\Delta$ is a partial function of the form $\Delta: (Q \times \Sigma \, \cup\, Q\times (\Sigma\times\oalph)) \to Q$, namely, $\Delta$ can be viewed as a partial function from its first coordinates to the last one. 
Note that every deterministic \rt is always unambiguous. 
The definition of unambiguous is in line with the notion of unambiguous annotated grammar~\cite{AmarilliJMR22} (see also \cite{MunozR22}), and determinism with the idea of I/O-determinism used in~\cite{FlorenzanoRUVV20,BourhisGJR21,grez2021formal}.
As usual, one can easily show that for every \rt $\cA$ there exists an equivalent deterministic \rt (of exponential size) and, therefore, an equivalent unambiguous \rt (see~\cite{FlorenzanoRUVV20,BourhisGJR21,grez2021formal} for a proof of this result). 
\begin{lem}\label{ra:det}
	For every \rtname $\cA$ there exists a deterministic \rtname $\cA'$ such that $\sem{\cA}(d) = \sem{\cA'}(d)$ for every $d\in\Sigma^*$.
\end{lem}
%\begin{proof}
%	We use the standard determinization scheme for regular automata. Namely, let $\cA = (Q, \Sigma, \oalph, \Delta, \qinit, F)$, and let $\cA' = (Q', \Sigma, \oalph, \Delta', \{\qinit\}, F')$ where $Q' = 2^Q$, $\Delta'$ is made of states $(S,b,T)$ such that $S,T\subseteq Q$, $b\in\Sigma\cup(\Sigma\times\oalph)$ and $T$ is the set of reachable states $q$ from some $p\in S$ by a transition $(p,b,q)\in\Delta$, and $F' = \{S\mid S\cap F\neq\emptyset\}$. It can be seen that $\cA'$ is deterministic. To show that $\sem{\cA}(d) = \sem{\cA'}(d)$ for every document $d\in\Sigma^*$, we define the {\it language} of an annotated automaton $\cA$ as a set $L(\cA)\subseteq (\Sigma \cup (\Sigma\cup\oalph))^*$ as follows:
%	$$
%		L(\cA) = \{b_1\ldots b_n\mid \text{There exists an accepting run $\rho = q_1 \xrightarrow{b_1} q_2 \xrightarrow{b_2} \ldots \xrightarrow{b_{n}} q_{n+1}$ of $\cA$}\}
%	$$
%	It is straightforward to see that $L(\cA) = L(\cA')$, which directly implies that for each document $d\in\Sigma^*$ it holds that $\sem{\cA}(d) = \sem{\cA'}(d)$.
%\end{proof}

\newcommand{\blabla}{Proposition 1 }

Regarding the expressive power of annotated automata, we note that they have the same expressive power as MSO formulas with monadic second-order free variables. We refer the reader to~\cite{MunozR22} for an analogous result in the context of nested documents. In fact, the equivalence between these models can be obtained as a corollary of \blabla in~\cite{MunozR22}.
Finally, by Lemma~\ref{ra:det} we do not lose expressive power if we restrict to the class unambiguous annotated automata, since we can convert every annotated automata into an deterministic one with an exponential cost in the size of the initial automaton. 

\paragraph{Main result} We are interested in the problem of evaluating annotated automata over an SLP-compressed document, namely, to enumerate all the annotations over the document represented by an SLP. Formally, we define the main evaluation problem of this paper as follows. Let $\mathcal{C}$ be any class of \rt (e.g., unambiguous \rt). 

\vspace{0.5em}

\begin{center}
	\framebox{
		\begin{tabular}{rl}
			\textbf{Problem:} \!\!\!\!\!\!& $\enumraa[\mathcal{C}]$  \\ \hline \vspace{-3mm} \\
			\textbf{Preprocessing:}  \!\!\!\!\!\! & 
			$\left\{ \text{\begin{tabular}{rl}
					\textbf{input:} & \!\!\! An \rt $\cA \in \mathcal{C}$ and an SLP $S$ \\
					%                                \textbf{output:}
					\textbf{result:}
					& \!\!\! A data structure $\DS_{\cA,S}$
			\end{tabular}} \right.
			$            \\ \hline \vspace{-3mm} \\
			\textbf{Enumeration:} \!\!\!\!\!\! &
			$\left\{
			\!\!\text{
				\begin{tabular}{rl}
					\textbf{input:} & \!\!\!
					\parbox[t]{4.2cm}{The data structure $\DS_{\cA,S}$} \\
					\textbf{output:} & \!\!\! \parbox[t]{4.2cm}{Enumerate $\sem{\cA}(\doc(S))$}
				\end{tabular} 
			}\right.$\!\!\!\!
		\end{tabular}
	}
\end{center}

\vspace{0.5em}
As is common for enumeration problems, we explicitly divide the problem $\enumraa[\mathcal{C}]$ in two phases: the \emph{preprocessing} and the \emph{enumeration}.  
First, the preprocessing phase receives $\cA$ and $S$ and produces some data structure $\DS_{\cA,S}$ (e.g., a collection of indices) which encodes the expected output. 
After the preprocessing it comes the enumeration phase that receives as input the data structure  $\DS_{\cA,S}$ and enumerates the results.
We say that such an algorithm has \emph{$f$-preprocessing time} if there exists a constant $c$ such that, for every input $(\cA, S)$, the time for the preprocessing phase of $(\cA, S)$ is bounded by $c \cdot f(\cA, S)$.   
Also, we say that the algorithm has \emph{output-linear delay} if there exists a constant $d$ such that whenever the enumeration phase delivers all outputs $O_1, \ldots, O_\ell$ sequentially, the time for producing the next output $O_i$ is bounded by $d \cdot |O_i|$ for every $i \leq \ell$.

%As it is common for enumeration problems, we want to impose an efficiency guarantee regarding the preprocessing of the input (in this case, $\cA$ and $S$) and the delay between two consecutive outputs. For this purpose, one often divides the work of the enumeration algorithm into two phases: first, a \emph{preprocessing phase} in which it receives the input and produces some data structure $D$ (e.g., a collection of indices) which encodes the expected output and, second, an \emph{enumeration phase} which extracts the results from $D$. 
%We say that such an algorithm has \emph{$f$-preprocessing time} if there exists a constant $c$ such that, for every input $\cI$, the time for the preprocessing phase of $\cI$ is bounded by $c \cdot f(|\cI|)$.   
%Instead, we say that the algorithm has \emph{output-linear delay} if there exists a constant $d$ such that whenever the enumeration phase delivers all outputs $O_1, \ldots, O_\ell$ sequentially, the time for producing the next output $O_i$ is bounded by $c \cdot |O_i|$ for every $i \leq \ell$.

As it is common in the literature~\cite{Segoufin13}, we assume here the computational model of Random Access Machines (RAM) with uniform cost measure and addition and subtraction as basic operations~\cite{aho1974design}. 
Further, as it is commonly done on algorithms over SLPs and other compression schemes, we assume that the registers in the underlying RAM-model allow for constant-time arithmetical operations over positions in the {\it uncompressed} document (i.e., they have $\cO(\log |\doc(S)|)$ size).

The notion of output-linear delay is a refinement of the better-known constant-delay bound, which requires that each output has a constant size (i.e., with respect to the input). Since even the document encoded by an SLP can be of exponential length, it is more reasonable in our setting to use the output-linear delay guarantee.

The following is the main technical result of this work.

\begin{thm}\label{theo:main-theorem}
	Let $\mathcal{C}$ be the class of all unambiguous \rts. Then one can solve the problem $\enumraa[\mathcal{C}]$ with linear preprocessing time and output-linear delay. Specifically, there exists an enumeration algorithm that runs in $|\cA|^3 \times |S|$-preprocessing time and output-linear delay for enumerating $\br{\cA}(\doc(S))$ given an unambiguous \rt $\cA$ and an SLP $S$. 
\end{thm}

We dedicate Sections~\ref{sec:ecs}--\ref{sec:evaluation} to presenting the enumeration algorithm of Theorem~\ref{theo:main-theorem}. In Section~\ref{sec:evaluation} we explain the preprocessing phase of the algorithm. Before that (Sections~\ref{sec:ecs} and~\ref{sec:ecs-op}), we explain how \emph{Enumerable Compact Sets with Shifts} work, which is the data structure that we use to store the outputs during the preprocessing phase.

%
%
%In this paper we will not only look at enumeration algorithms as a solution to a two-part problem, instead we will also be interested in accepting these objects $D$ as part of an input.
%Another consideration we will have is that we will only talk about output-linear delay as our gold standard for enumeration efficiency. 
%This is a refinement of the better-known constant-delay bound, which requires (by a reasonable assumption) that each output has constant size. 
%Output-linear delay is the natural extension of this notion which allows outputs of unbounded size, which requires the delay to be bounded linearly in the size of the current output.
%More generally, we will look at classes of objects which allow output-linear delay enumeration as schemes which efficiently represent sets of outputs.
%We can subsequently define enumeration algorithms as those which receive an input, and produce one such scheme.

	\section{Enumerable compact sets with shifts}\label{sec:ecs}
	
	%!TEX root = ../main.tex

In this section, we present the data structure, called \emph{Enumerable Compact Sets with Shifts} (\dsabbr), that will be used to compactly store the outputs of evaluating an annotated automaton over a straight-line program. This structure extends Enumerable Compact Sets (ECS) introduced in~\cite{MunozR22} (which were, in turn, strongly inspired by the work in~\cite{AmarilliBJM17,AmarilliBMN19}). Indeed, people have also used ECS extensions in~\cite{AmarilliJMR22,BucchiGQRV22,RiverosJV23}. This new version extends ECS by introducing a shift operator, which allows to succinctly move all outputs' positions with a single call. Although the shift nodes require a revision of the complete ECS model, it simplifies the evaluation algorithm in Section~\ref{sec:evaluation} and achieves output-linear delay for enumerating all the outputs. For completeness of presentation, this section goes through all the formalization and enumeration algorithm for \dsabbr, as it was done in~\cite{MunozR22}, and how to modify them with shifts. Afterwards, in Section~\ref{sec:ecs-op}, we present the operations to modify \dsabbr that will be crucial in the evaluation algorithm in Section~\ref{sec:evaluation}.
%We will follow a preprint of the full version of~\cite{MunozR22} as base of this section~\cite{MunozR22Arxiv}.

\paragraph{The structure} Next, we first provide the definition of a structure and the semantics of our data structure, and then we give an example. The reader may find it helpful to look at Figure~\ref{fig-ecsex} when reading the following definitions. 

Let $\infAlph$ be an output alphabet such that $\infAlph$ has no elements in common with $\int$ or $\{\cup,\odot\}$ (i.e., $\infAlph \cap \int = \emptyset$ and $\infAlph \cap \{\cup,\odot\} = \emptyset$).
We define an \emph{\dsnamebigcaps{}} (\dsabbr) as a tuple:\[
\cD \ = \  (\infAlph, V, \lch, \rch, \lambda)
\] 
such that $V$ is a finite sets of nodes,
$\lch\colon V \to V$ and $\rch\colon V \to V$ are the {\em left} and {\em right} partial functions, and $\lambda\colon V\to\infAlph \cup \int \cup\{\cup,\odot\}$ is a labeling function.
We call $\ell(v)$ and $r(v)$ the \emph{children} of $v$.
Intuitively, a node defines a set of outputs, and its label indicates how they are formed: that is, whether it utilizes the outputs from its children, and how.
We assume that $\cD$ forms a directed acyclic graph, namely, the induced graph $(V, \{(v,\lch(v)),(v,\rch(v))\mid v\in V\})$ is acyclic. 
Further, for every node $v \in V$, $\lch(v)$ is defined iff  $\lambda(v)\in \int \cup \{\cup, \odot\}$, and $\rch(v)$ is defined iff $\lambda(v)\in \{\cup, \odot\}$.
%Also, we will assume that if $\ell(v) = u$, then $\lambda(u)\not\in\int$.
% This last assumption implies that chains of $\int$-nodes are not allowed.
Note that, by definition, nodes labeled by $\infAlph$ are bottom nodes in the acyclic structure formed by $\cD$, and nodes labeled by $\int$ or $\{\cup, \odot\}$ are inner nodes. Here, $\int$-nodes are unary operators (i.e., $\rch(\cdot)$ is not defined over them), and $\cup$-nodes or $\odot$-nodes are binary operators. 
Indeed, we say that $v \in V$ is a \textdef{bottom node} if $\lambda(v) \in \infAlph$, a \textdef{product node} if $\lambda(v) = \odot$, a \textdef{union node} if $\lambda(v) = \cup$, and a \textdef{shift node} if $\lambda(v)\in\int$. 
Furthermore, we say that a node is an \textdef{output node} if it is a bottom or product node. 
Finally, we define the size of $\cD$ as $|\cD| = |V|$.

The outputs retrieved from a \dsabbr are strings of the form $(\oout_1, i_1)(\oout_2, i_2)\ldots(\oout_\ell, i_\ell)$, where $\oout_j\in\infAlph$ and $i_j\in\int$.
To build them, we use the {\em shifting function} $\shift: (\infAlph \times \int) \times \int \to (\infAlph \times \int)$ such that $\shift((\oout, i), s) = (\oout, i+s)$.
We extend this function to strings over $\infAlph \times \int$ such that $\shift((\oout_1, i_1)\ldots(\oout_\ell, i_\ell), s) = (\oout_1, i_1+s)\ldots(\oout_\ell, i_\ell+s)$ and to set of strings such that $\shift(L, s) = \{\shift(w, s) \mid w \in L\}$ for every $L \subseteq (\infAlph \times \int)^*$.

As we mentioned, each node $v \in V$ of a \dsabbr $\cD = (\infAlph, V, \lch, \rch, \lambda)$ defines a set of output strings. Specifically, we associate a set of strings $\sem{\cD}(v)$ recursively as follows. For any two sets of strings $L_1$ and $L_2$, define $L_1 \cdot L_2 = \{w_1\cdot w_2 \mid w_1\in L_1\text{ and }w_2\in L_2\}$. Then: 
\begin{itemize}
	\item if $\lambda(v) = \oout \in \infAlph$, then $\sem{\cD}(v) = \{(\oout, 1)\}$;
	\item if $\lambda(v) = \cup$, then $\sem{\cD}(v) = \sem{\cD}(\lch(v)) \cup \sem{\cD}(\rch(v))$;
	\item if $\lambda(v) = \odot$, then $\sem{\cD}(v) = \sem{\cD}(\lch(v)) \cdot \sem{\cD}(\rch(v))$; and 
	\item if $\lambda(v) \in \int$, then $\sem{\cD}(v) = \shift(\sem{\cD}(\lch(v)), \lambda(v))$.
\end{itemize}

\begin{exa}\label{exa:fig}
Suppose $\infAlph = \{x, y\}$. 
Consider the \dsabbr $\cD = (\infAlph, V, \lch, \rch, \lambda)$ where $V = \{v_1, v_2, v_3, v_4, v_5\}$, $\lch(v_1) = v_4$, $\rch(v_1) = v_2$, $\lch(v_2) = v_3$, $\lch(v_3) = v_4$, $\rch(v_3) = v_5$, $\lambda(v_1) = \odot$, $\lambda(v_2) = +2$, $\lambda(v_3) = \cup$, $\lambda(v_4) = x$ and $\lambda(v_5) = y$. 
We show an illustration of this \dsabbr~in~Figure~\ref{fig-ecsex}. One can easily check that the sets of words $\sem{\cD}$ associated to each node~are: 
\[
\begin{array}{rcl}
	\sem{\cD}(v_4) & = & \{(x, 1)\} \\  \sem{\cD}(v_5) & = & \{(y, 1)\} \\ \sem{\cD}(v_3) & = & \{(x, 1), (y, 1)\} \\ \sem{\cD}(v_2) & = & \{(x, 3), (y, 3)\} \\
	\sem{\cD}(v_1) & = & \{(x, 1)(x, 3), (x, 1)(y, 3)\}.
\end{array}
\]
\end{exa}
% \cristian{Ahora que tenemos espacio, yo traería la figura 3(a) a esta página y la removería de la figura 3. La figura 3 quedará un poco desbalanceada, no será tan terrible, pero ganaremos que para el lector es más fácil de verificar el ejemplo.}

% \begin{figure}[t]
% 	\centering
% 	\begin{tikzpicture}[every label/.style={black!60}, ->,>=stealth',roundnode/.style={circle,inner sep=1pt},squarednode/.style={rectangle,inner sep=2pt}, scale=0.85]

% 		\node [roundnode, label=180:\footnotesize{$v_4\!:$}] (4) at (0, 0) {$x$};
% 		\node [roundnode, label=180:\footnotesize{$v_5\!:$}] (5) at (4, 0) {$y$};
% 		\node [squarednode, label=180:\footnotesize{$v_3\!:$}] (3) at (2, 1) {$\cup$};
% 		\node [squarednode, label=180:\footnotesize{$v_2\!:$}] (2) at (3, 2) {$+2\ $};
% 		\node [squarednode, label=180:\footnotesize{$v_1\!:$}] (1) at (2, 3) {$\odot$};
% 		\draw[dashed] (1) to[out=-135,in=90] (4);
% 		\draw (1) to[out=-45,in=120] (2);
% 		\draw[dashed] (2) to[out=-135,in=45] (3);
% 		\draw[dashed] (3) to[out=-135,in=45] (4);
% 		\draw (3) to[out=-45,in=135] (5);
		
% 		\node at (2, -0.7) {(a)}; 

% 	\end{tikzpicture}
% 	\caption{(a) An example of a \dsabbr with output alphabet $\{x, y\}$. 
% 	% (b) Gadget for $\prod(\cD,v_1,v_2, k)$. (c) Gadget for $\union(\cD,v_3,v_4)$. We use dashed and solid edges for the left and right mappings, respectively. Node names are in grey at the left of each node. In (b) and (c), square nodes are the input and output nodes of each~operation.
% 	}
% 	\label{fig:gadgets}
% 	\vspace{-3mm}
% \end{figure}

\begin{figure}[t]
	\centering
	\begin{tikzpicture}[every label/.style={black!60}, ->,>=stealth',roundnode/.style={circle,inner sep=2pt},squarednode/.style={rectangle,inner sep=2pt}]
		\node [roundnode, label=180:$v_4\!:$] (4) at (0, 0) {$x$};
		\node [roundnode, label=180:$v_5\!:$] (5) at (4, 0) {$y$};
		\node [squarednode, label=180:$v_3\!:$] (3) at (2, 1) {$\cup$};
		\node [squarednode, label=180:$v_2\!:$] (2) at (3, 2) {$+2\ $};
		\node [squarednode, label=180:$v_1\!:$] (1) at (2, 3) {$\odot$};
		\draw[dashed] (1) to[out=-135,in=90] (4);
		\draw (1) to[out=-45,in=120] (2);
		\draw[dashed] (2) to[out=-135,in=45] (3);
		\draw[dashed] (3) to[out=-135,in=45] (4);
		\draw (3) to[out=-45,in=135] (5);
	\end{tikzpicture}
	\caption{An example of an \dsabbr with output alphabet $\{x, y\}$ where $v_1$ is a product node, $v_2$ is a shift node, $v_3$ is a union node, and $v_4$ and $v_5$ are bottom nodes. We use dashed and solid edges for the left  and right partial function, respectively.} 
	\label{fig-ecsex}
\end{figure}
\paragraph{The enumeration algorithm} Given that every node of a \dsabbr represents a set of strings, we are interested in enumerating them with output-linear delay. Specifically, we focus on the following problem. Let $\mathcal{C}$ be a class of \dsnamesbigcaps{}.

% \vspace{0.2em}

\begin{center}
	\framebox{
		\begin{tabular}{rl}
			\textbf{Problem:} & $\enumecs[\mathcal{C}]$\\
			\textbf{Input:} & A \dsabbr $\cD \in \mathcal{C}$ and a node $v$ of $\cD$ \\
			\textbf{Output:} & Enumerate $\sem{\cD}(v)$.
		\end{tabular}
	}
\end{center}

\vspace{0.2em}

The plan is then to provide an enumeration algorithm with output-linear delay for the problem $\enumecs[\mathcal{C}]$ and some helpful class $\mathcal{C}$.
A reasonable strategy to enumerate the set $\sem{\cD}(v)$ is to do a traversal on the structure while accumulating the shift values in the path to each leaf. However, to be able to do this without repetitions and output-linear delay, we need to guarantee two conditions: first, that one can obtain every output from $\cD$ in only one way and, second, union and shift nodes are \emph{close} to an output node,
%(i.e., a bottom node or a product node), 
in the sense that we can always reach them in a bounded number of steps. To ensure that these conditions hold, we impose two restrictions on an ECS. 

\begin{enumerate}
	\item[(i)] We say that $\D$ is \textdef{duplicate-free} if $\D$ satisfies the following two properties: (1) for every union node $v$ it holds that $\sem{\D}(\ell(v))$ and $\sem{\D}(r(v))$ are disjoint, and (2) for every product node $v$ and for every $w \in \sem{\D}(v)$, there exists a unique way to decompose $w = w_1 \cdot w_2$ such that $w_1 \in \sem{\D}(\ell(v))$ and $w_2 \in \sem{\D}(r(v))$. Intuitively, this condition is what ensures the enumeration can be done with no repeats.
	
	\item[(ii)] We define the notion of \emph{$k$-bounded}~\dsabbr{} as follows. 
	Given an \dsabbr{} $\D$, define the {\em (left) output-depth} of a node $v\in V$, denoted by $\odepth_{\D}(v)$, recursively as follows:
	$\odepth_{\D}(v) = 0$ whenever $\lambda(v)\in\infAlph$ or $\lambda(v) = \odot$ (i.e., $v$ is an output node), and $\odepth_{\D}(v) = \odepth_{\D}(\ell(v))+1$ whenever $\lambda(v) \in \{\cup\}\cup\int$.
	Then, for $k\in\nat$ we say that $\D$ is $k$-bounded if $\odepth_{\D}(v)\leq k$ for all $v\in V$. Intuitively, we only limit the distance to the first output node we see to the left. Since this condition holds for every node, this is enough to ensure that in the enumeration phase the next output string $w$ can always be retrieved in $|w|$ time.
\end{enumerate}

%For the first restriction, we say that an ECS $\cD$ is \emph{duplicate-free} if the following hold: (1) for every union node $v$ in $\cD$ it holds that $\sem{\cD}(\lch(v))$ and $\sem{\cD}(\rch(v))$ are disjoint and (2) for every product node $v$ and for every $w \in \sem{\cD}(v)$, there exists a unique way to decompose $w = w_1 \cdot w_2$ such that $w_1 \in \sem{\cD}(\lch(v))$ and $w_2 \in \sem{\cD}(\rch(v))$.
%
%For the second restriction, we define \emph{$k$-bounded}~\dsabbr{}. 
%Given a \dsabbr{} $\cD$, define the (left) output-depth of a node $v\in V$, denoted by $\odepth_{\cD}(v)$, recursively as follows:
%$\odepth_{\cD}(v) = 0$ whenever $\lambda(v)\in \{\odot\}\cup\Omega$, and $\odepth_{\cD}(v) = \odepth_{\cD}(\lch(v))+1$ whenever $\lambda(v) \in \{\cup\}\cup\int$.
%Then, for $k\in\nat$ we say that $\cD$ is $k$-bounded if $\odepth_{\cD}(v)\leq k$ for all $v\in V$.

\begin{prop}\label{prop:lindelay}
	Fix $k\in\nat$. Let $\mathcal{C}_k$ be the class of all  duplicate-free and $k$-bounded \dsabbrs{}. Then one can solve the problem $\enumecs[\mathcal{C}_k]$ with output-linear delay and constant preprocessing time (or equivalently, without preprocessing).
\end{prop}
\begin{proof}
	\newcommand{\len}{\mathsf{len}}
	
	Let $\D = (\infAlph, V, \ell, r, \lambda)$ be a duplicate-free and $k$-bounded \dsabbr{}.
	The algorithm that we present is a depth-first traversal of the DAG, done in a recursive fashion to ensure that after retrieving some output $w$, the next one $w'$ can be printed in $O(k\cdot (|w| + |w'|))$ time. The entire procedure is detailed in Algorithm~\ref{alg:enumeration}.
	%!TEX root = ../main/main.tex

\algdef{SE}[DOWHILE]{Do}{doWhile}{\algorithmicdo}[1]{\algorithmicwhile\ #1}%
\algnewcommand{\LineComment}[1]{\State \(\triangleright\) #1}

\begin{algorithm*}[t]
	\caption{Enumeration over a node $u$ from some ECS $\cD = (\infAlph, V, \ell, r, \lambda)$.}\label{alg:enumeration}
	\smallskip
	\begin{varwidth}[t]{0.5\textwidth}
		\begin{algorithmic}[1]
			
			\Procedure{{enumerate}}{$v$} \label{alg1enum1}
			\State $\uit \gets \textsc{create}(v)$
			\While{$\uit.\textsc{next} = \textbf{true}$}\label{alg1enum2}
			\State $\uit.\textsc{print}(0)$\label{alg1enum3}
			\EndWhile
			\EndProcedure
			
			\State
			
			\LineComment{Bottom node iterator $\uit_\infAlph$} \label{alg1bottom0}
			\Procedure{{create}}{$v$} 
			\State \Comment{Assuming\ $\lambda(v) \in \infAlph$} \label{alg1bottom1}
			\State $\bnodelabel \gets v$ \label{alg1bottom2}
			\State $\hasnext \gets \textbf{true}$ \label{alg1bottom3}
			\EndProcedure	
			
			\State
			
			\Procedure{{next}}{} \label{alg1bottom4}
			\If{$\hasnext = \textbf{true}$} \label{alg1bottom5}
			\State $\hasnext \gets \textbf{false}$ \label{alg1bottom6}
			\State \Return $\textbf{true}$ \label{alg1bottom7}
			\EndIf
			\State \Return $\textbf{false}$ \label{alg1bottom8}
			\EndProcedure
			
			\State
			
			\Procedure{{print}}{$s$} \label{alg1bottom9}
			\State $\textbf{print}: (\lambda(\bnodelabel), 1+s)$ \label{alg1bottom10}
			\EndProcedure \label{alg1bottom11}
			
			\State
			
			\LineComment{Product node iterator $\uit_\odot$} \label{alg1prod0}
			\Procedure{{create}}{$v$}
			\State \Comment{Assuming\ $\lambda(v) = \odot$}\label{alg1prod1}
			\State $u \gets v$
			\State $\leftit \gets \textsc{create}(\ell(u))$
			\State $\leftit.\textsc{next}$
			\State $\rightit \gets \textsc{create}(r(u))$\label{alg1prod5}
			\EndProcedure	
			
			\State
			
			\Procedure{{next}}{} \label{alg1prod6}
			\If{$\rightit.\textsc{next} = \textbf{false}$}
			\If{$\leftit.\textsc{next} = \textbf{false}$} \label{alg1prod8}
			\State \Return $\textbf{false}$
			\EndIf
			\State $\rightit \gets \textsc{create}(r(u))$
			\State $\rightit.\textsc{next}$ \label{alg1prod11}
			\EndIf
			\State \Return $\textbf{true}$ \label{alg1prod12}
			\EndProcedure
			
			\State
			
			\Procedure{{print}}{$s$} \label{alg1prod13}
			\State $\leftit.\textsc{print}(s)$
			\State $\rightit.\textsc{print}(s)$ \label{alg1prodlast}
			\EndProcedure
			\algstore{myalg}
		\end{algorithmic}
	\end{varwidth} \hspace{4.5em}
	\begin{varwidth}[t]{0.5\textwidth}
		\begin{algorithmic}[1]
			\algrestore{myalg}
			\LineComment{Union/$\int$ node iterator $\uit_{\cup/\int}$} \label{alg1union0}
			\Procedure{{create}}{$v$}\label{alg1union1}
			\State \Comment{Assuming $\lambda(v) \in \{\cup\}\cup\int$}
			\State $\ustack\gets\text{\sf push}(\ustack, (v, 0)\,)$\label{alg1union4}
			\State $\ustack\gets\textsc{traverse}(\ustack)$\label{alg1union5}
			\State $\uit \gets \textsc{create}(\text{\sf top}(\ustack).u)$ \label{alg1union6}
			\EndProcedure	
			
			\State
			
			\Procedure{{next}}{}
			\If{$\uit.\textsc{next} = \textbf{false}$}\label{alg1unionnext1}
			\State $\ustack\gets\text{\sf pop}(\ustack)$ \label{alg1unionnext2}
			\If{$\text{\sf depth}(\ustack) = 0$} \label{alg1unionnext3}
			\State \Return $\textbf{false}$
			\EndIf
			\State $(u,s) \gets \text{\sf top}(\ustack)$ \label{alg1unionnext5}
			\If{$\lambda(u) \in \{\cup\} \cup \int$} 
			\State $\ustack\gets\textsc{traverse}(\ustack)$ \label{alg1unionnext6}
			\EndIf
			\State $\uit \gets \textsc{create}(u)$ \label{alg1unionnext7}
			\State $\uit.\textsc{next}$ \label{alg1unionnext8}
			\EndIf
			\State \Return $\textbf{true}$
			\EndProcedure
			
			\State
			
			\Procedure{{print}}{$s$} 
			\State $(u, s')\gets \text{\sf top}(\ustack)$ \label{alg1unionprint1}
			\State $\uit.\textsc{print}(s + s')$ \label{alg1unionprint2}
			\EndProcedure
			
			\State
			
			\Procedure{{traverse}}{$\ustack$}\label{alg1traverse0}
			\While{$\lambda(\text{\sf top}(\ustack).u) \in \{\cup\} \cup \int$}
			\State $(u,s) \gets \text{\sf top}(\ustack)$
			\State $\ustack \gets \text{\sf pop}(\ustack)$
			\If{$\lambda(u)\in\int$} \label{alg1traverseIfStart}
			\State $v\gets \ell(u)$
			\State $s'\gets s + \lambda(u)$
			\State $\ustack\gets\text{\sf push}(\ustack, (v, \, s'))$ \label{alg1traverseIfEnd}
			\Else \label{alg1traverseElseStart}
			\State $\ustack\gets\text{\sf push}(\ustack, (r(u), s)\,)$
			\State $\ustack\gets\text{\sf push}(\ustack, (\ell(u),s)\,)$ \label{alg1traverseElseEnd}
			\EndIf
			\EndWhile
			\State \Return $\ustack$ \label{alg1unionlast}
			\EndProcedure
		\end{algorithmic}
	\end{varwidth}\hspace{1.8em}
	\smallskip
\end{algorithm*}

	To simplify the presentation of the algorithm, we use an \emph{iterator interface} that, given a node $v$, contains all information and methods to enumerate the outputs $\sem{\D}(v)$. Specifically, an iterator $\uit$ must implement the following three methods:
	\[
	\begin{array}{rclrclrcl}
		\textsc{create}(v) & \!\!\!\!\rightarrow\!\!\!\! & \uit   \ \ \ \ \ \	 & \uit.\textsc{next} & \!\!\!\! \rightarrow \!\!\!\! & b  \ \ \ \ \ \ \ &  \uit.\textsc{print}(s)  & \!\!\!\!\rightarrow\!\!\!\! & \emptyset
	\end{array}
	\]
	where $v$ is a node, $b$ is either {\bf true} or {\bf false}, and $\emptyset$ means that the method does not return an output. 
	The first method, \textsc{create}, receives a node $v$ and creates an iterator $\uit$ of the type of $v$. We will implement three types of iterators, one for bottom nodes ($\uit_\infAlph$), one for product nodes ($\uit_\odot$), and one for union and $\int$-nodes together ($\uit_{\cup/\int}$). The second method, $\uit.\textsc{next}$, moves the iterator to the next output, returning {\bf true} if, and only if, there is an output to print. Then the last method, $\uit.\textsc{print}$, receives an integer value $s$, and writes the current output pointed by $\uit$ to the output registers after shifting the output by $s$. 
	We assume that, after creating an iterator $\uit$, one must first call $\uit.\textsc{next}$ to move to the first output before printing. Furthermore, if $\uit.\textsc{next}$ outputs {\bf false}, then the behavior of $\uit.\textsc{print}$ is undefined. Note that one can call $\uit.\textsc{print}$ several times, without calling $\uit.\textsc{next}$, and the iterator will write the same output each time in the output registers. 
	
	Assume we can implement the iterator interface for each type. Then the procedure $\textsc{Enumerate}(v)$ in Algorithm~\ref{alg:enumeration} (lines \ref{alg1enum1}-\ref{alg1enum3}) shows how to enumerate the set $\sem{\D}(v)$ by using an iterator $\uit$ for $v$. In the following, we show how to implement the iterator interface for each type and how the size of the next output bounds the delay between two outputs.

	We start by presenting the iterator $\uit_\infAlph$ for a bottom node $v$ (lines \ref{alg1bottom0}-\ref{alg1bottom11}), called a \emph{bottom node iterator}. We assume that each $\uit_\infAlph$ has internally two values, denoted by $\bnodelabel$ and $\hasnext$, where $\bnodelabel$ is a reference to $v$ and $\hasnext$ is a boolean variable. The purpose of a bottom node iterator is only to print $(\lambda(u), 1+s)$ for some shift $s$. For this goal, when we create $\uit_\infAlph$, we initialize $u$ equal to $v$ and $\hasnext = \textbf{true}$ (lines \ref{alg1bottom2}-\ref{alg1bottom3}). Then, when we call $\uit_\infAlph.\textsc{next}$ for the first time, we swap $\hasnext$ from {\bf true} to {\bf false} and output {\bf true} (i.e., there is one output ready to be returned). Then any following call to $\uit_\infAlph.\textsc{next}$ will be false (lines \ref{alg1bottom4}-\ref{alg1bottom8}). Finally, the $\uit_\infAlph.\textsc{print}$ writes the pair $(\lambda(u),1+s)$ to the output registers (lines \ref{alg1bottom9}-\ref{alg1bottom11}). Here, we assume the existence of a method {\bf print} on the RAM model for writing the next entry to the output~registers.

	For a product node, we present a \emph{product node iterator} $\uit_\odot$ in Algorithm~\ref{alg:enumeration} (lines \ref{alg1prod0}-\ref{alg1prodlast}). 
	This iterator receives a product node $v$ with $\lambda(v) = \odot$ and stores a reference of $v$, called $u$, and two iterators $\leftit$ and $\rightit$, for iterating through the left and right node, $\ell(u)$ and $r(u)$, respectively. 
	The \textsc{create} method initializes $u$ with $v$, creates the iterators $\leftit$ and $\rightit$, and calls $\leftit.\textsc{next}$ to be prepared for the first call of $\uit_\odot.\textsc{next}$ (lines \ref{alg1prod1}-\ref{alg1prod5}).
	The purpose of $\uit_\odot.\textsc{next}$ is to fix one output for the left node $\ell(u)$ and iterate over all outputs of $r(u)$ (lines \ref{alg1prod6}-\ref{alg1prod12}). When we stop enumerating all outputs of $\sem{\D}(r(u))$, we move to the next output of $\leftit$, and iterate again over all $\sem{\D}(r(u))$ (lines \ref{alg1prod8}-\ref{alg1prod11}).
	For printing, we recursively call first the printing method of $\leftit$, and then the one of $\rightit$ (lines \ref{alg1prod13}-\ref{alg1prodlast}).
		
	The most involved case is the \emph{union/$\int$ node iterator} $\uit_{\cup/\int}$ (lines \ref{alg1union0}-\ref{alg1unionlast}). 
	This iterator receives a node $v$ of one of two types, either a union node, or a $\int$-node. It keeps a \emph{stack} $\ustack$ and an iterator~$\uit$. The elements in the stack are pairs $(u, s)$ where $u$ is a node and $s$ is an integer. We assume the standard implementation of a stack with the native methods $\text{\sf push}$, $\text{\sf pop}$, $\text{\sf top}$, and $\text{\sf depth}$: the first three define the standard operations over stacks, and $\text{\sf depth}$ counts the elements in a stack. 
	The purpose of the stack is to perform a \emph{depth-first-search} traversal of all union and $\int$ nodes below $v$, reaching all possible output nodes $u$ such that there is a path of only union and $\int$ nodes between $v$ and $u$. At every point, if an element $(u,s)$ is in the stack, then $s$ is equal to the sum of all $\int$ nodes in the path from $v$ to $u$. If the top node of $\ustack$ is a pair $(u, s)$ such that $u$ is an output node, then $\uit$ is an iterator for $u$, which enumerates all their outputs. If $p = (u,s)$, we will use the notation $p.u$ to access the first component of $p$ (i.e., $u$).
	
	In order to perform the \emph{depth-first-search} traversal of union and $\int$ nodes, we use the auxiliary method $\textsc{traverse}(\ustack)$ (lines \ref{alg1traverse0}-\ref{alg1unionlast}). While the node $u$ at the top of $\ustack$ is a union or a shift node, we pop the top pair $(u,s)$ from $\ustack$. If $u$ is a shift node (lines~\ref{alg1traverseIfStart}-\ref{alg1traverseIfEnd}), we push the pair $(v, s')$ in the stack where $v$ is the node $\ell(u)$ pointed by $u$ and $s'$ is the sum of the current shift $s$ with the shift $\lambda(u)$.
	Otherwise, if $u$ is a union node (lines~\ref{alg1traverseElseStart}-\ref{alg1traverseElseEnd}), we first push the right pair $(r(u), s)$ followed by the left pair $(\ell(u), s)$ into the stack. 	
	The while-loop will eventually reach an output node at the top of the stack and end. 
	It is important to note that $\textsc{traverse}(\ustack)$ takes $\cO(k)$ steps, given that the ECS is $k$-bounded. Then if $k$ is fixed, the  $\textsc{traverse}$ procedure takes constant time. In Figure~\ref{fig-enum-stacks}, we illustrate the evolution of a stack $\ustack$ inside a union node iterator when we call $\textsc{traverse}(\ustack)$ several times. 
	\begin{figure}[t]
		\centering
		%!TEX root = ../main/main.tex
\newcommand{\figfourscaleval}{0.83}
\begin{tikzpicture}[scale=\figfourscaleval,->,>=stealth',roundnode/.style={circle,draw,inner sep=1.2pt},squarednode/.style={rectangle,inner sep=3pt}]
	
	\node [squarednode] (num) at (0, 5.5) {1.};
	\node [squarednode] (0) at (1, 0) {$a_3$};
	%	\node [roundnode] (0b) at (1, 0.3) {};
	\node [squarednode] (1) at (3, 0) {$a_4$};
	%	\node [roundnode] (1b) at (3, 0.3) {};
	\node [squarednode] (2) at (2, 1) {$\cup$};
	\node [squarednode, inner sep=1pt] (2p) at (1.7, 1.85) {\scriptsize \raisebox{0.2em}{$+2$}};
	%	\node [roundnode] (2b) at (2, 1.3) {};
	\node [squarednode] (3) at (0, 1.5) {$a_2$};
	%	\node [roundnode] (3b) at (0, 1.3) {};
	\node [squarednode] (4) at (1, 2.5) {$\cup$};
	%	\node [roundnode] (4b) at (1, 2.3) {};
	\node [squarednode] (5) at (3, 2.5) {$a_5$};
	%	\node [roundnode] (5b) at (3, 2.3) {};
	\node [squarednode] (6) at (2, 3.5) {$\cup$};
	\node [squarednode, inner sep=1pt] (6p) at (1.6, 4.35) {\scriptsize \raisebox{0.2em}{$+3$}};
	%	\node [roundnode] (6b) at (2, 3.3) {};
	\node [squarednode] (7) at (0, 4) {$a_1$};
	\node [squarednode] (8) at (1, 5) {$\cup$};
	\node [squarednode] (9) at (3, 5) {$a_6$};
	%	\node [roundnode] (9b) at (3, 4.3) {};
	\node [squarednode] (10) at (2, 6) {$\cup$};
	\node [squarednode, left=-1pt of 10] (10l) {$v:$};
	%	\node [roundnode] (12b) at (1, 6.3) {};
	\node [squarednode] () at (0, -1) {$\ustack : $};
	\node [squarednode] () at (2, -1) {\tiny $(\cup, 0)$};
	
	\draw (2) to (0);
	\draw (2) to (1);
	\draw[-] (4) to (2p);
	\draw[out=-90] (2p) to (2);
	\draw (4) to (3);
	\draw (6) to (4);
	\draw (6) to (5);
	\draw[-] (8) to (6p);
	\draw[out=-90] (6p) to (6);
	\draw (8) to (7);
	\draw (10) to (8);
	\draw (10) to (9);
	
	\draw[dashed] (0.5,6) to (10l);
	
\end{tikzpicture}
\hspace{1em}
\begin{tikzpicture}[scale=\figfourscaleval,->,>=stealth',roundnode/.style={circle,draw,inner sep=1.2pt},squarednode/.style={rectangle,inner sep=3pt}]
	
	\node [squarednode] (num) at (0, 5.5) {2.};
\node [squarednode] (0) at (1, 0) {$a_3$};
%	\node [roundnode] (0b) at (1, 0.3) {};
\node [squarednode] (1) at (3, 0) {$a_4$};
%	\node [roundnode] (1b) at (3, 0.3) {};
\node [squarednode] (2) at (2, 1) {$\cup$};
\node [squarednode, inner sep=1pt] (2p) at (1.7, 1.85) {\scriptsize \raisebox{0.2em}{$+2$}};
%	\node [roundnode] (2b) at (2, 1.3) {};
\node [squarednode] (3) at (0, 1.5) {$a_2$};
%	\node [roundnode] (3b) at (0, 1.3) {};
\node [squarednode] (4) at (1, 2.5) {$\cup$};
%	\node [roundnode] (4b) at (1, 2.3) {};
\node [squarednode] (5) at (3, 2.5) {$a_5$};
%	\node [roundnode] (5b) at (3, 2.3) {};
\node [squarednode] (6) at (2, 3.5) {$\cup$};
\node [squarednode, inner sep=3pt] (6p) at (1.6, 4.35) {\scriptsize \raisebox{0.2em}{$+3$}};
%	\node [roundnode] (6b) at (2, 3.3) {};
\node [squarednode] (7) at (0, 4) {$a_1$};
\node [squarednode] (8) at (1, 5) {$\cup$};
\node [squarednode] (9) at (3, 5) {$a_6$};
%	\node [roundnode] (9b) at (3, 4.3) {};
\node [squarednode] (10) at (2, 6) {$\cup$};
\node [squarednode, left=-1pt of 10] (10l) {$v:$};
%	\node [roundnode] (12b) at (1, 6.3) {};
\node [squarednode] () at (1.5, -1) {\tiny $(a_1,0)(+3,0)(a_6, 0)$};

\draw (2) to (0);
\draw (2) to (1);
\draw[-] (4) to (2p);
\draw[out=-90] (2p) to (2);
\draw (4) to (3);
\draw (6) to (4);
\draw (6) to (5);
\draw[-] (8) to (6p);
\draw[out=-90] (6p) to (6);
\draw (8) to (7);
\draw (10) to (8);
\draw (10) to (9);

	\draw [dashed] (3,5.7) to (9);
	\draw [dashed,out=-90,in=0] (9) to (6p);
	\draw [dashed,out=-180,in=0] (6p) to (7);
	
\end{tikzpicture}
\hspace{1em}
\begin{tikzpicture}[scale=\figfourscaleval,->,>=stealth',roundnode/.style={circle,draw,inner sep=1.2pt},squarednode/.style={rectangle,inner sep=3pt}]
	
	\node [squarednode] (num) at (0, 5.5) {3.};
	\node [squarednode] (0) at (1, 0) {$a_3$};
	%	\node [roundnode] (0b) at (1, 0.3) {};
	\node [squarednode] (1) at (3, 0) {$a_4$};
	%	\node [roundnode] (1b) at (3, 0.3) {};
	\node [squarednode] (2) at (2, 1) {$\cup$};
	\node [squarednode, inner sep=3pt] (2p) at (1.7, 1.85) {\scriptsize \raisebox{0.2em}{$+2$}};
	%	\node [roundnode] (2b) at (2, 1.3) {};
	\node [squarednode] (3) at (0, 1.5) {$a_2$};
	%	\node [roundnode] (3b) at (0, 1.3) {};
	\node [squarednode] (4) at (1, 2.5) {$\cup$};
	%	\node [roundnode] (4b) at (1, 2.3) {};
	\node [squarednode] (5) at (3, 2.5) {$a_5$};
	%	\node [roundnode] (5b) at (3, 2.3) {};
	\node [squarednode] (6) at (2, 3.5) {$\cup$};
	\node [squarednode, inner sep=1pt] (6p) at (1.6, 4.35) {\scriptsize \raisebox{0.2em}{$+3$}};
	%	\node [roundnode] (6b) at (2, 3.3) {};
	\node [squarednode] (7) at (0, 4) {$a_1$};
	\node [squarednode] (8) at (1, 5) {$\cup$};
	\node [squarednode] (9) at (3, 5) {$a_6$};
	%	\node [roundnode] (9b) at (3, 4.3) {};
	\node [squarednode] (10) at (2, 6) {$\cup$};
	\node [squarednode, left=-1pt of 10] (10l) {$v:$};
	%	\node [roundnode] (12b) at (1, 6.3) {};
	\node [squarednode] () at (1.5, -1) {\tiny $(a_2,3)(+2,3)(a_5,3)(a_6, 0)$};
	
	\draw (2) to (0);
	\draw (2) to (1);
	\draw[-] (4) to (2p);
	\draw[out=-90] (2p) to (2);
	\draw (4) to (3);
	\draw (6) to (4);
	\draw (6) to (5);
	\draw[-] (8) to (6p);
	\draw[out=-90] (6p) to (6);
	\draw (8) to (7);
	\draw (10) to (8);
	\draw (10) to (9);
	
	\draw [dashed] (3,5.7) to (9);
	\draw [dashed] (9) to (5);
	\draw [dashed,out=-90,in=0] (5) to (2p);
	\draw [dashed,out=-180,in=0] (2p) to (3);
	
\end{tikzpicture}
\hspace{1em}
\begin{tikzpicture}[scale=\figfourscaleval,->,>=stealth',roundnode/.style={circle,draw,inner sep=1.2pt},squarednode/.style={rectangle,inner sep=3pt}]
	
\node [squarednode] (num) at (0, 5.5) {4.};
\node [squarednode] (0) at (1, 0) {$a_3$};
%	\node [roundnode] (0b) at (1, 0.3) {};
\node [squarednode] (1) at (3, 0) {$a_4$};
%	\node [roundnode] (1b) at (3, 0.3) {};
\node [squarednode] (2) at (2, 1) {$\cup$};
\node [squarednode, inner sep=1pt] (2p) at (1.7, 1.85) {\scriptsize \raisebox{0.2em}{$+2$}};
%	\node [roundnode] (2b) at (2, 1.3) {};
\node [squarednode] (3) at (0, 1.5) {$a_2$};
%	\node [roundnode] (3b) at (0, 1.3) {};
\node [squarednode] (4) at (1, 2.5) {$\cup$};
%	\node [roundnode] (4b) at (1, 2.3) {};
\node [squarednode] (5) at (3, 2.5) {$a_5$};
%	\node [roundnode] (5b) at (3, 2.3) {};
\node [squarednode] (6) at (2, 3.5) {$\cup$};
\node [squarednode, inner sep=1pt] (6p) at (1.6, 4.35) {\scriptsize \raisebox{0.2em}{$+3$}};
%	\node [roundnode] (6b) at (2, 3.3) {};
\node [squarednode] (7) at (0, 4) {$a_1$};
\node [squarednode] (8) at (1, 5) {$\cup$};
\node [squarednode] (9) at (3, 5) {$a_6$};
%	\node [roundnode] (9b) at (3, 4.3) {};
\node [squarednode] (10) at (2, 6) {$\cup$};
\node [squarednode, left=-1pt of 10] (10l) {$v:$};
%	\node [roundnode] (12b) at (1, 6.3) {};
\node [squarednode] () at (1.5, -1) {\tiny $(a_3,5)(a_4,5)(a_5,3)(a_6,0)$};

\draw (2) to (0);
\draw (2) to (1);
\draw[-] (4) to (2p);
\draw[out=-90] (2p) to (2);
\draw (4) to (3);
\draw (6) to (4);
\draw (6) to (5);
\draw[-] (8) to (6p);
\draw[out=-90] (6p) to (6);
\draw (8) to (7);
\draw (10) to (8);
\draw (10) to (9);

	\draw [dashed] (3,5.7) to (9);
	\draw [dashed] (9) to (5);
	\draw [dashed,out=-90,in=45] (5) to (1);
	\draw [dashed] (1) to (0);
	
\end{tikzpicture}
		\caption{Evolution of the stack $\ustack$ (written on the bottom and represented by dashed arrows) for an iterator over the node $v$ in the figure. 
		The underlying ECS is made of union nodes, two $\int$ nodes, and six bottom nodes. The first figure is $\ustack$ after calling $\ustack \gets {\sf push}(\ustack, (v,0))$, the second is after calling $\ustack \gets\textsc{Traverse}(\ustack)$. The last two figures represent successive calls to ${\sf pop}(\ustack), \ustack \gets\textsc{Traverse}(\ustack)$.}
		% \cristian{TODO: actualizar la figura y leyenda según la nueva versión del traverse.}
		\label{fig-enum-stacks}
	\end{figure}
	
	The methods of a union/$\int$ node iterator $\uit_{\cup/\int}$ are then straightforward. For \textsc{create} (lines \ref{alg1union1}-\ref{alg1union6}), we push $(v,0)$ and then traverse $\ustack$, finding the first leftmost output node from~$v$~(lines~\ref{alg1union4}-\ref{alg1union5}). 
	Then we build the iterator $\uit$ of this output node for being ready to start enumerating their outputs (line~\ref{alg1union6}). 
	For \textsc{next}, we consume all outputs by calling $\uit.\textsc{next}$ (line~\ref{alg1unionnext1}). When there are no more outputs, we pop the top node from $\ustack$ and check if the stack is empty or not (lines~\ref{alg1unionnext2}-\ref{alg1unionnext3}). If this is the case, there are no more outputs and we output {\bf false}. Otherwise, if $\ustack$ is non-empty but the top pair $(u,s)$ of $\ustack$ contains a union node, then we apply the \textsc{traverse} method for finding the leftmost output node from~$u$~(lines \ref{alg1unionnext5}-\ref{alg1unionnext6}). When the procedure is done, we know that the node in the top pair is an output node, and then we create an iterator and move to its first output (lines~\ref{alg1unionnext7}-\ref{alg1unionnext8}). 
	For $\textsc{print}(s)$, we see the pair $(u, s')$ at the top, where we recall that $s'$ represents the sum of all $\int$ nodes on the way to $u$ (line~\ref{alg1unionprint1}), and $u$ is assumed to be an output node. Then, we call the print method of $\uit$ which is ready to write the current output, and over which we add the value $s+s'$ (line~\ref{alg1unionprint2}). 
		
	In order to prove the correctness of the enumeration procedure, one can verify that $\textsc{Enumerate}(v)$ in Algorithm~\ref{alg:enumeration} enumerates all the outputs in the set $\sem{\D}(v)$ one by one, and without repetitions, which follows from the fact that $\D$ is duplicate-free. To bound the delay between outputs, the fact that $\D$ is $k$-bounded implies that the delay is bounded by $\cO(k\cdot |w_0|)$ if $w_0$ is the first output, or $O(k\cdot (|w| + |w'|))$ if $w$ and $w'$ are the previous and next output, respectively. More precisely, we note that every iterator $\tau$ defines a {\em current} output $w_{\sf curr}$: the string that is printed when calling $\tau.{\textsc{print}}(s)$; and a {\em next} output $w_{\sf next}$: the string that is printed when calling first $\tau.\textsc{next}$ and then $\tau.\textsc{print}$; which give us following runtime bounds:
	\begin{itemize}
		\item $\textsc{create}(v)$ takes time $\cO(k\cdot |w_0|)$, where $w_0$ is the first output in $\sem{\D}(v)$,
		\item $\textsc{next}$ takes time $\cO(k\cdot |w_0|)$ for the first call, and $\cO(k\cdot (|w_{\sf curr}|+|w_{\sf next}|))$ on each subsequent call,
		\item $\textsc{print}(s)$ takes time $\cO(k\cdot |w_{\sf curr}|)$.
	\end{itemize}
	Overall, $\textsc{Enumerate}(v)$ in Algorithm~\ref{alg:enumeration} requires $O(k\cdot (|w| + |w'|))$ delay to write the next output $w'$ in the output register, after printing the previous output $w$. 
	
	We end by pointing out that the existence of an enumeration algorithm $\enumE$ with delay $O(k\cdot (|w| + |w'|))$ between any consecutive outputs $w$ and $w'$, implies the existence of an enumeration algorithm $\enumE'$ with output-linear delay as defined in Section~\ref{sec:setting}. We start noting that $k$ is a fixed value and then the delay of $\enumE$ only depends on $|w| + |w'|$. For depending only on the next output $w'$, one can perform the following strategy for $\enumE'$: start by running~$\enumE$, enumerate the first output $w_0$, advance $k \cdot |w_0|$ more steps of $\enumE$, and stop. Then continue running $\enumE$, enumerate the next output $w_1$, advance $k \cdot |w_1|$ more steps, and stop\footnote{If advancing $k \cdot |w_1|$ more steps requires printing part of the next outputs $w_2, w_3, \ldots$, we could store these outputs in some temporary registers of the RAM model to retrieve them later.}. By repeating this enumeration process, one can verify that the delay between the $i$-th output $w_i$ and the $(i+1)$-th output $w_{i+1}$ is $O(|w_{i+1}|)$. Therefore, $\enumE'$ has output-linear delay. 
\end{proof}

	\section{Operations over Shift-ECS}\label{sec:ecs-op}
	
	%!TEX root = ../main.tex

The previous section introduces the structure of a \dsabbr{} and how to enumerate its outputs. In this section, we provide the operations we allow to extend an \dsabbr{}, maintaining the good properties for enumeration (i.e., $k$-boundedness). We start by presenting the semantics and implementations of these operations, to then introduce two special nodes that will be instrumental for our evaluation algorithm.

%The next step is to provide a set of operations that allow extending a \dsabbr{} $\cD$ in a way that maintains $k$-boundedness. 
\paragraph{The operations} Fix a \dsabbr{} $\cD = (\infAlph, V, \lch, \rch, \lambda)$. Then for any $\oout \in \infAlph$, $v_1, \ldots, v_4, v \in V$ and $s\in\int$, we define these four operations, given alongside their expected semantics:
%\[
%\begin{array}{rclrclrcl}
%	\add(\oout) \! \! \! &\to&\! \! \! v'  \ \ \ \ \ \  \prod(v_1, v_2) \! \! \! &\to&\! \! \! v'   \\  \union(v_3, v_4) \! \! \! &\to&\! \! \! v' \ \ \ \ \ \  \shiftop(v, k) \! \! \! \! \! &\to&\! \! \! v'
%\end{array}
%\]
\[
\begin{array}{rl}
	\add(\oout) \ \to\  v'  \ \ \ \ \ \  &\sem{\cD}(v') := \{(\oout,1)\}\\
	\prod(v_1, v_2) \ \to\ v' \ \ \ \ \ \  &\sem{\cD}(v') := \sem{\cD}(v_1) \cdot \sem{\cD}(v_2) \\  
	\union(v_3, v_4) \ \to\  v' \ \ \ \ \ \ &\sem{\cD}(v') := \sem{\cD}(v_3) \cup \sem{\cD}(v_4)\\
	\shiftop(v, s) \ \to\ v' \ \ \ \ \ \ &\sem{\cD}(v') := \shift(\sem{\cD}(v), s)
\end{array}
\]
Intuitively, $\add$ allows to add a new bottom node, and $\prod$, $\union$, and $\shiftop$ to do the product, union, and shift of nodes, respectively.
Here we assume that the $\union$ and $\prod$ are called only when properties (1) and (2) of a duplicate-free \dsabbr{} hold, namely, $\sem{\cD}(v_3)$ and $\sem{\cD}(v_4)$ are disjoint; and, for every $w \in \sem{\cD}(v_1) \cdot \sem{\cD}(v_2)$, there exists a unique way to decompose $w = w_1 \cdot w_2$ such that $w_1 \in \sem{\cD}(v_1)$ and $w_2 \in \sem{\cD}(v_2)$.

Strictly speaking, each operation above should receive as input the data structure $\cD$, and output a fresh node $v'$ plus a new data structure $\cD' = (\infAlph, V', \lch', \rch', \lambda')$ such that $\cD'$ is an extension of $\cD$, namely,
that $V\subseteq V'$, and for each $v\in V$ it holds that $\ell'(v) = \ell(v)$, $r'(v) = r(v)$ and $\lambda'(v) = \lambda(v)$. Note that we assume that each operation would only extend the data structure with new nodes and that old nodes are immutable after each operation. By this, we mean that once a node $v$ is created, and the values $\ell(v)$, $r(v)$ and $\lambda(v)$ assigned, none of these values ever change again. This also implies that the entire subgraph that is reachable from $v$, and thus, the set $\sem{\cD}(v)$, also stay the same. For simplification, we will not explicitly refer to $\cD$ on the operations above, although they modify $\cD$ directly by adding new nodes. 

\paragraph{Implementation of the operations} To implement the above operations, we impose further restrictions on the structure below the operations' input nodes to ensure $k$-boundedness. Towards this goal, we introduce the notion of \textdef{safe nodes}.
We say that a node $v \in V$ is \emph{safe} if it is a shift node over an output node, or over a union node with an output node on the left and a shift node with output depth of at most 2 on the right. More precisely, $v$ is safe iff $\lambda(v) \in \mathbb{Z}$, and either: $\lch(v)$ is an output node; or $\lch(v)$ is a union node with $\odepth_{\cD}(\lch(v)) = 1$ and $\rch(\lch(v))$ is a shift node with $\odepth_{\cD}(\rch(\lch(v))) \leq 2$.
In other words, $v$ is safe if 
The next step is to show that all operations over \dsabbrs{} receive only safe nodes and output only safe nodes. 
As we will see, safeness will be a light structural restriction on the operations' input nodes that is enough to maintain $k$-boundedness after each operation.  

Next, we show how to implement each operation assuming that every input node is safe. In fact, the cases of $\add$ and $\shiftop$ are straightforward. 
For $\add(\oout)\to v'$ we extend $\cD$ with two fresh nodes $v'$ and $u$ such that $\lambda(u) = \oout$, $\lambda(v') = 0$, and $\lch(v') = u$. In other words, we hang a fresh $0$-shift node $v'$ over a fresh $\oout$-node $u$, and output $v'$. 
For $\shiftop(v, s) \to v'$, add the fresh node $v'$ to $\cD$, and set $\ell(v') = \ell(v)$ and $\lambda(v') = \lambda(v) + s$.
One can easily check that in both cases the node $v'$ represents the desired set, is safe, and $k$-boundedness is preserved.

To show how to implement $\prod(v_1, v_2)\to v'$, recall that $v_1$ and $v_2$ are safe and, in particular, both are shift nodes. 
Then we need to extend $\cD$ with fresh nodes $v'$, $v''$, and $v'''$ such that  $\ell(v') = v''$, $\ell(v'') = \ell(v_1)$, $r(v'') = v'''$, $\ell(v''') = \ell(v_2)$, $\lambda(v') = \lambda(v_1)$, $\lambda(v'') = \odot$ and $\lambda(v''') = \lambda(v_2) - \lambda(v_1)$. 
Figure~\ref{fig:gadgets}(a) shows a diagram of this gadget. One can easily check that $v'$ represents the product of $v_1$ and $v_2$, $v'$ is safe, and the new version of $\cD$ is $k$-bounded whenever $\cD$ is also $k$-bounded.

The last operation is $\union(v_3, v_4)\to v'$. The strategy is then to prove that if $v_3$ and $v_4$ are safe nodes, then we can implement the operator and produce a safe node $v'$. Let us define $v'$ as follows:
\begin{itemize}
	\item If at least one among $\ell(v_3)$ and $\ell(v_4)$ is an output node, assume without loss of generality that it is $\ell(v_3)$. We extend $\cD$ with nodes $v'$, $v''$ and $v'''$, where $\ell(v') = v''$, $\ell(v'') = \ell(v_3)$, $r(v'') = v'''$, $\ell(v''') = \ell(v_4)$, $\lambda(v') = \lambda(v_3)$, $\lambda(v'') = \cup$ and $\lambda(v''') = \lambda(v_4) - \lambda(v_3)$. 
	This construction is identical to the $\prod$ construction shown in Figure~\ref{fig:gadgets}(a), except replacing $v_1$ and $v_2$ by $v_3$ and $v_4$ respectively, and $\lambda(v'')$ from $\odot$ to $\cup$.
	%This construction is illustrated in FIgure~\ref{fig-theo-ds}. 
%	\item When both $u_3 = \lch(v_3)$ and $u_4 = \lch(v_4)$ are union nodes, let $k'_1 = \lambda(r(u_3))$ and let $k'_2 = \lambda(r(u_4))$. 
%	We extend $\cD$ with fresh nodes $v',v'_1,v'_2,v'_3,v'_4,v'_5$ and $v'_6$. 
%	Define $\lambda(v') = k_1$, $\lambda(v'_1) = \cup$, $\lambda(v'_2) = k_2 - k_1$, $\lambda(v'_3) = \cup$,  $\lambda(v'_4) = k_1+k'_1-k_2$, $\lambda(v'_5) = \cup$ and $\lambda(v'_6) = k_2 + k'_2 - k_1 - k'_1$. 
%	Then define $\ell(v') = v'_1$, $\ell(v'_1) = \ell(u_3)$, $r(v'_1) = v'_2$, $\ell(v'_2) = v'_3$, $\ell(v'_3) = \ell(u_4)$, $r(v'_3) = v'_4$, $\ell(v'_4) = v'_5$, $\ell(v'_5) = \ell(r(u_3))$, $r(v'_5) = v'_6$ and $\ell(v'_6) = \ell(r(u_4))$. 
	\item When both $u_3 = \lch(v_3)$ and $u_4 = \lch(v_4)$ are union nodes, let $s'_1 = \lambda(r(u_3))$ and let $s'_2 = \lambda(r(u_4))$. 
	We extend $\cD$ with fresh nodes $v',v'_1,v'_2,v'_3,v'_4,v'_5$ and $v'_6$. 
	Define $\lambda(v') = s_1$, $\lambda(v'_1) = \cup$, $\lambda(v'_2) = s_2 - s_1$, $\lambda(v'_3) = \cup$,  $\lambda(v'_4) = s_1+s'_1-s_2$, $\lambda(v'_5) = \cup$ and $\lambda(v'_6) = s_2 + s'_2 - s_1 - s'_1$. 
	Then define $\ell(v') = v'_1$, $\ell(v'_1) = \ell(u_3)$, $r(v'_1) = v'_2$, $\ell(v'_2) = v'_3$, $\ell(v'_3) = \ell(u_4)$, $r(v'_3) = v'_4$, $\ell(v'_4) = v'_5$, $\ell(v'_5) = \ell(r(u_3))$, $r(v'_5) = v'_6$ and $\ell(v'_6) = \ell(r(u_4))$. 
	We show an illustration of this gadget in Figure~\ref{fig:gadgets}(b).
\end{itemize}

We can prove that the last construction has several interesting properties.
First, one can check that $\sem{\cD}(v') = \sem{\cD}(v_3)\cup \sem{\cD}(v_4)$ since each shift value is 
%carefully 
constructed so that the accumulated shift value from $v'$ to each node remains unchanged.
Thus, the semantics is well-defined. 
Second, $\union$ can be computed in constant time in $|\cD|$ given that we only need to add a fixed number of fresh nodes.
%and the operation is fully-persistent given that we connect them to previous nodes without modifying $\cD$. 
Furthermore, the produced node $v'$ is safe, even though some of the new nodes are not necessarily safe. 
Finally, the new $\cD$ is 3-bounded whenever $\cD$ is 3-bounded. This is straightforward to see for the case when $\ell(v_3)$ (or $\ell(v_4)$) is an output node. To see this for the second case, we first have to notice that $\ell(u_3)$ and $\ell(u_4)$ are output nodes, and that $\odepth(\ell(r(u_3)))\leq 1$ and $\odepth(\ell(r(u_4)))\leq 1$. We can check the depth of each node going from the bottom to the top: $\odepth(v_6') \leq 2$, $\odepth(v_5') \leq 2$, $\odepth(v_4') \leq 3$, $\odepth(v_3')\leq 1$, $\odepth(v_2') \leq 2$, $\odepth(v_1') \leq 1$ and $\odepth(v')\leq 2$.

\begin{figure}[t]
	\centering
	\begin{tikzpicture}[every label/.style={black!60}, ->,>=stealth',roundnode/.style={circle,inner sep=1pt},squarednode/.style={rectangle,inner sep=2pt}, scale=0.85]
		
		% \begin{scope}[xshift=-7cm, yshift=6cm]
		% \node [roundnode, label=180:\footnotesize{$v_4\!:$}] (4) at (0, 0) {$x$};
		% \node [roundnode, label=180:\footnotesize{$v_5\!:$}] (5) at (4, 0) {$y$};
		% \node [squarednode, label=180:\footnotesize{$v_3\!:$}] (3) at (2, 1) {$\cup$};
		% \node [squarednode, label=180:\footnotesize{$v_2\!:$}] (2) at (3, 2) {$+2\ $};
		% \node [squarednode, label=180:\footnotesize{$v_1\!:$}] (1) at (2, 3) {$\odot$};
		% \draw[dashed] (1) to[out=-135,in=90] (4);
		% \draw (1) to[out=-45,in=120] (2);
		% \draw[dashed] (2) to[out=-135,in=45] (3);
		% \draw[dashed] (3) to[out=-135,in=45] (4);
		% \draw (3) to[out=-45,in=135] (5);
		
		% \node at (2, -0.7) {(a)}; 
		% \end{scope}
	
		\begin{scope}[xshift=-8cm, yshift=0cm]
			\node[squarednode] (0) at (0, 0) {$\ell(v_1)$};
			\node[squarednode, draw, label=180:\footnotesize{$v_1\!:$}] (2) at (2, 1.2) {\raisebox{-0.55em}{$s_1$}};
			\node[squarednode] (3) at (4, 0) {$\ell(v_2)$};
			\node[squarednode, draw, label=180:\footnotesize{$v_2\!:$}] (5) at (6, 1.2) {\raisebox{-0.55em}{$s_2$}};
			\node[squarednode, label=180:\footnotesize{$v'''\!:$}] (7) at (5.5, 2.1) {$(s_2-s_1)$};
			\node[squarednode, label=180:\footnotesize{$v''\!:$}] (8) at (3, 3.0) {$\odot$};
			\node[squarednode, draw, label=180:\footnotesize{$v'\!:$}] (9) at (4.5, 4.2) {\raisebox{-0.55em}{$s_1$}};
			\draw[dashed] (2.south west) to[out=-135,in=45] (0);
			\draw[dashed] (5.south west) to[out=-135,in=45] (3);
			\draw[dashed] (7.south west) to[out=-135,in=90] (3);
			\draw (8) to[out=-45,in=135] (7.north west);
			\draw[dashed] (8) to[out=-135,in=90] (0);
			\draw[dashed] (9.south west) to[out=-135,in=90] (8);
			
			\node at (3, -0.7) {(a)}; 
	
		\end{scope}

		\node[squarednode] (0) at (0, 0) {$\ell(u_3)$};
		\node[squarednode] (1) at (2, 0) {$\ell(r(u_3))$};
		\node[squarednode, label=180:\footnotesize{$r(u_3)\!:$}] (1p) at (2, 1) {$s_1'$};
		\node[squarednode, label=180:\footnotesize{$u_3\!:$}] (2) at (1, 2) {$\cup$};
		\node[squarednode, draw, label=180:\footnotesize{$v_3\!:$}] (2p) at (1.2, 3) {\raisebox{-0.55em}{$s_1$}};
		\node[squarednode] (0b) at (4, 0) {$\ell(u_4)$};
		\node[squarednode] (1b) at (6, 0) {$\ell(r(u_4))$};
		\node[squarednode, label=180:\footnotesize{$r(u_4)\!:$}] (1bp) at (6, 1) {$s_2'$};
		\node[squarednode, label=180:\footnotesize{$u_4\!:$}] (2b) at (5, 2) {$\cup$};
		\node[squarednode, draw, label=180:\footnotesize{$v_4\!:$}] (2bp) at (5.2, 3) {\raisebox{-0.55em}{$s_2$}};
		\node[squarednode, draw, label=180:\footnotesize{$v'\!:$}] (vp) at (3.5, 9) {\raisebox{-0.55em}{$s_1$}};
		\node[squarednode, label=180:\footnotesize{$v_1'\!:$}] (v) at (3, 8) {$\cup$};
		\node[squarednode, label=180:\footnotesize{$v_2'\!:$}] (u1p) at (4.5, 7) {$s_2 - s_1$};
		\node[squarednode, label=180:\footnotesize{$v_3'\!:$}] (u1) at (4, 6) {$\cup$};
		\node[squarednode, label=180:\footnotesize{$v_4'\!:$}] (u2p) at (5.5, 5) {$s_1+s_1'-s_2$};
		\node[squarednode, label=180:\footnotesize{$v_5'\!:$}] (u2) at (4.8, 4) {$\cup$};
		\node[squarednode, label=180:\footnotesize{$v_6'\!:\!\!\!\!\!\!\!$}] (u3) at (7, 3) {$
			\begin{array}{r}
			s_2 +s_2' \\
			- s_1 - s_1'
			\end{array}$};
		\draw[dashed] (vp) to (v);
		\draw[dashed] (v) to[out=-135, in=110] (0);
		\draw (v) to[out=-45] (u1p);
		\draw[dashed] (u1p) to (u1);
		\draw[dashed] (u1) to[out=-135, in=110] (0b);
		\draw (u1) to[out=-45] (u2p);
		\draw[dashed] (u2p) to (u2);
		\draw[dashed] (u2) to[out=-135, in=45] (1);
		\draw (u2) to[out=-45] (u3);
		\draw[dashed] (u3) to[out=-135,in=45] (1b);
		\draw[dashed] (2) to[out=-135, in=90] (0);
		\draw (2) to[out=-45,in=135] (1p);
		\draw[dashed] (1p) to[out=-135,in=90] (1);
		\draw[dashed] (2p) to (2);
		\draw[dashed] (2b) to[out=-135, in=90] (0b);
		\draw (2b) to[out=-45,in=135] (1bp);
		\draw[dashed] (1bp) to[out=-135,in=90] (1b);
		\draw[dashed] (2bp) to (2b);
		
		\node at (3, -0.7) {(b)}; 
	\end{tikzpicture}
	\caption{
	(a) Gadget for $\prod(v_1,v_2)$. 
	(b) Gadget for $\union(v_3,v_4)$. 
	We use dashed and solid edges for the left and right child, respectively. 
	Node names are in grey at the left of each node. Nodes in square boxes are the input and output nodes of each~operation.}
	\label{fig:gadgets}
%	\vspace{-3mm}
\end{figure}

By the previous discussion, if we start with a \dsabbr\ $\cD$ which is 3-bounded (in particular, empty) and we apply the $\add$, $\prod$, $\union$ and $\shiftop$ operators between safe nodes (which also produce safe nodes), then the result is 3-bounded as well.
Furthermore, the data structure is fully-persistent~\cite{DriscollSST86}: for every node $v$ in $\cD$, $\sem{\cD}(v)$ is immutable after each operation. Finally, by Proposition~\ref{prop:lindelay}, the result can be enumerated with output-linear~delay. In other words, we have the following result.

\begin{prop}\label{prop:data-structure}
	The operations $\add$, $\prod$, $\union$ and $\shiftop$ take constant time and are fully persistent. Furthermore, if we start from an empty \dsabbr\ $\cD$ and apply these operations over safe nodes, the result node $v'$ is always a safe node and the set $\sem{\cD}(v)$ can be enumerated with output-linear delay (without preprocessing) for every node $v$.
\end{prop} 

We summarized the main ideas of the operations over a \dsabbr{} in the previous proposition. Next, we show how to extend Proposition~\ref{prop:data-structure} (i.e., Theorem~\ref{theo:data-structure-eps} below) to include two special nodes that will be instrumental for running our enumeration algorithm.

\paragraph{The empty- and $\epsilon$-nodes} The last step of constructing our model of \dsabbr{} is the inclusion of two special nodes that produce the empty set and the empty string, called empty- and $\epsilon$-node, respectively. 
%For the sake of presentation, we include both nodes here, although they are crucial for our evaluation algorithm in Section~\ref{sec:evaluation}.

We start with the \emph{empty node}, which is easier to incorporate into a \dsabbr{}. Consider a special node $\bot$ and include it on every \dsabbr{} $\cD$, such that $\sem{\cD}(\bot) = \emptyset$. 
Then extend the operations $\prod$, $\union$, and $\shiftop$ accordingly to the empty set, namely, $\prod(v, \bot) = \prod(\bot, v) \to \bot$, $\union(v, \bot) = \union(\bot, v) \to v$, and $\shiftop(\bot, s) \to \bot$ for every node $v$, and $s \in \int$. One can easily include the $\bot$-node into \dsabbrs{} without affecting the guarantees of Proposition~\ref{prop:data-structure}: The operator $\add$ is unchanged since $\bot$ is always part of the \dsabbr{}, the operator $\prod$ and $\shift$ can be hardcoded to output $\bot$ if an argument is ever $\bot$, and the operator $\union$ can output $v$ in the same scenario. None of the nodes in $\cD$ ever reach $\bot$ so the enumeration phase sees no change.

The other special node is the \emph{$\epsilon$-node}. Let $\eps$ denote a special node, included on every \dsabbr{} $\cD$, such that $\sem{\cD}(\epsilon) = \{\epsilon\}$. With these new nodes in a \dsabbr{}, we need to revise our notions of output-depth, duplicate-free, and $k$-boundedness to change the enumeration algorithm, and to extend the operations $\add$, $\prod$, $\union$, and $\shiftop$ over so-called $\epsilon$-safe nodes (i.e., the extension of safe nodes with $\epsilon$). 

 Our overall strategy will be to greatly restrict the places the $\eps$-node can appear. This will essentially separate the node from the rest of the structure where it will minimally affect the enumeration phase. We will also show how to rewrite the operations $\add$, $\prod$, $\union$ and $\shiftop$ so that their semantics are fully adapted to having the $\eps$-node while maintaining the aforementioned restriction. 

%Given space restrictions, we show the use of $\epsilon$-nodes in an extended version and how we can preserve Proposition~\ref{prop:lindelay} and Theorem~\ref{theo:data-structure}. 

%	Extending these four operations to deal with empty nodes is straightforward, and this has already been defined, so we will focus on operations that deal with the $\eps$-node.

The first move will be to assume every \dsabbr{} $\cD$ satisfies what we call the \emph{$\eps$-condition}: that is, the only nodes that can have the $\eps$-node as a child must be (1) union nodes, and (2) parentless. Further, if the $\eps$-node is a child, it must be the left child of its parent.
	
	The notions of output-depth, duplicate-free, and $k$-boundedness are mostly unchanged in this setting: If $v$ is the $\eps$-node, then $\odepth(v) = 0$, and if $v$ is a union node with the $\eps$-node as its left child, $\odepth(v) = 1$. No other node has the $\eps$-node as a descendant, so the definition of output-depth remains the same; and the definitions of duplicate-free and $k$-bounded remain unchanged as well. By assuming the $\eps$-condition, one can see that enumeration remains identical as before, except now the $\eps$-node might be seen once, and always as either the root or the left child of the root. Thus, we can state the following as a corollary of Proposition~\ref{prop:lindelay}:
	
	\begin{cor}\label{prop:lindelayeps}
		Fix $k\in\nat$. Let $\mathcal{C}_k$ be the class of all  duplicate-free and $k$-bounded \dsabbrs{} that satisfy the $\eps$-condition. Then one can solve the problem $\enumecs[\mathcal{C}_k]$ with output-linear delay and without preprocessing (i.e. constant preprocessing time).
	\end{cor}

	We now define the notion of \textdef{$\eps$-safe} nodes. These are nodes $v$ which satisfy one of three conditions: (1) $v$ is the $\eps$-node, (2) $v$ is safe and none of its descendants is the $\eps$-node, and (3) $v$ is a union node, its left child is the $\eps$-node, and its right child is a safe node for which none of its descendants is the $\eps$-node. 
	To guide the constructions, we call these conditions 1, 2 and 3.
	These conditions will allow us to use the constructions already given for $\prod$, $\union$, and $\shiftop$ over safe nodes, such as node $v$, in the case of condition 2, or $r(v)$, in the case of condition 3.
	The rest of this proof will be extending these operations to work over $\eps$-safe nodes, and check that they can be done in constant time, return an $\eps$-safe node, maintain the $\eps$-condition in the \dsabbr, and satisfy the specifications of each operation (i.e., that for the resulting node $v'$ the set $\sem{\cD}(v')$ contains what it should). 
	
	We define the operation $\add$ the same as in a \dsabbr without $\eps$. The operation $\add(\eps)$ is not defined since the $\eps$-node is always part of $\cD$.
	
	The $\prod$ case is somewhat more involved. First, assume that some node among $v_1$ and $v_2$ is the $\eps$-node (condition 1). Without loss of generality, assume that $v_1 = \eps$ . If $v_2$ is the $\eps$-node as well, we simply return $v_1$; and if $v_2$ is not the $\eps$-node (conditions 2 or 3), we simply return $v_2$. The requirements of the construction follow trivially. Now, assume that both $v_1$ and $v_2$ satisfy condition 2. For this case we simply return $\prod(v_1, v_2)$ as it was defined for a \dsabbr without $\eps$. Lastly, assume at least one node among $v_1$ and $v_2$ satisfies condition~3 and none of them are the $\eps$-node. These cases are depicted in Figure~\ref{fig-prod-multi-gadget}. 
	We describe them formally as follows, assuming that $v'$ is the output and using the operations $\union$ and $\prod$ as they were defined for a \dsabbr without $\eps$.
	\begin{itemize}
		\item If $v_1$ satisfies condition 2 and $v_2$ satisfies condition 3, define $v'' \gets \prod( v_1, r(v_2))$ and $v' \gets \union( v_1, v'')$.
		\item If $v_1$ satisfies condition 3 and $v_2$ satisfies condition 2, define $v'' \gets \prod( r(v_1), v_2)$ and $v' \gets \union( v'', v_2)$.
		\item If both $v_1$ and $v_2$ satisfy condition 3, define $v'' \gets \prod( r(v_1), r(v_2))$, $v_3 \gets \union( v'', r(v_2))$, and $v_4 \gets \union( r(v_1), v_3)$. Lastly, define $\lambda(v') = \cup$, $\ell(v') = \eps$, and $r(v') = v_4$.
	\end{itemize}
	
	One can verify that these constructions work as expected, namely, $\sem{\cD'}(v) = \sem{\cD}(v_1)\cdot\sem{\cD}(v_2)$. Also, if $v_1$ and $v_2$ are $\eps$-safe, then $v'$ is $\eps$-safe as well. Also, each construction does a fixed number of steps, so they take constant time.
	
	\begin{figure}[t]
		\centering
		\hfill
		\begin{tikzpicture}[->,>=stealth',roundnode/.style={circle,draw,inner sep=1.2pt},squarednode/.style={rectangle,inner sep=2pt}, scale=1]
			\node [squarednode] (c) at (1.12, -0.6) {$(a)$};
			\node [squarednode] (0) at (0.8, 2.75) {$\union$};
			\node [squarednode] (6) at (-0.05, 2.75) {$v'\, \colon$};
			\node [squarednode] (1) at (0, 1) {$v_1$};
			\node [squarednode] (2) at (1.6, 1.75) {$\prod$};
			\node [squarednode] (3) at (1.6, 1) {$v_2$};
			\node [squarednode] (4) at (0.8, 0) {$\eps$};
			\node [squarednode] (5) at (2.4, 0) {$r(v_2)$};
			\draw[dashed] (0) to[out=-135,in=90] (1);
			\draw (0) to[out=-45,in=120] (2);
			\draw (2) to[out=-45,in=90] (5);
			\draw[dashed] (3) to[out=-135,in=60] (4);
			\draw (3) to[out=-45,in=120] (5);
			\draw[dashed] (2) to (1);
		\end{tikzpicture} \hfill
		\begin{tikzpicture}[->,>=stealth',roundnode/.style={circle,draw,inner sep=1.2pt},squarednode/.style={rectangle,inner sep=2pt}, scale=1]
			\node [squarednode] (c) at (1.6, -0.6) {$(b)$};
			\node [squarednode] (0) at (2.5, 2.75) {$\union$};
			\node [squarednode] (6) at (1.65, 2.75) {$v'\, \colon$};
			\node [squarednode] (1) at (3, 1) {$v_2$};
			\node [squarednode] (2) at (1.90, 1.75) {$\ \prod$};
			\node [squarednode] (3) at (0.85, 1) {$v_1$};
			\node [squarednode] (4) at (0.2, 0) {$\eps$};
			\node [squarednode] (5) at (1.5, 0) {$r(v_1)$};
			\draw (0) to[out=-60,in=90] (1);
			\draw[dashed] (0) to[out=-135,in=70] (2);
			\draw[dashed] (2) to[out=-120,in=90] (5);
			\draw[dashed] (3) to[out=-135,in=60] (4);
			\draw (3) to[out=-45,in=120] (5);
			\draw (2) to (1);
		\end{tikzpicture} \hfill
		\begin{tikzpicture}[->,>=stealth',roundnode/.style={circle,draw,inner sep=1.1pt},squarednode/.style={rectangle,inner sep=2pt},scale=1.1]
			\node [squarednode] (c) at (1.6, -0.5) {$(c)$};
			\node [squarednode] (0) at (1.5, 2.75) {$\cup$};
			\node [squarednode] (11) at (0.88, 2.8) {$v'\, \colon$};
			\node [squarednode] (1) at (0.75, 2.25) {$\eps$};
			\node [squarednode] (2) at (2.25, 2.25) {$\union$};
			\node [squarednode] (3) at (3, 1.75) {$\union$};
			\node [squarednode] (4) at (2.1, 1) {$\prod$};
			\node [squarednode] (5) at (3.2, 1) {$v_2$};
			\node [squarednode] (6) at (2.5, 0) {$\eps$};
			\node [squarednode] (7) at (4, 0) {$r(v_2)$};
			\node [squarednode] (8) at (0.5, 1) {$v_1$};
			\node [squarednode] (9) at (0, 0) {$\eps$};
			\node [squarednode] (10) at (1, 0) {$r(v_1)$};
			\draw[dashed] (0) to (1);
			\draw (0) to (2);
			\draw[dashed] (2) to[out=-135,in=70] (10);
			\draw[dashed] (4) to (10);
			\draw (4) to[out=-45,in=145] (7);
			\draw[dashed] (5) to[out=-135,in=60] (6);
			\draw (5) to[out=-45,in=120] (7);
			\draw (3) to[out=-45,in=90] (7);
			\draw[dashed] (3) to (4);
			\draw (2) to (3);
			\draw (8) to[out=-50,in=100] (10);
			\draw[dashed] (8) to[out=-125,in=80] (9);
		\end{tikzpicture} \ \ \hfill
		\caption{Gadgets for $\prod$ as defined for a \dsabbr with the $\eps$-node.}
		\label{fig-prod-multi-gadget}
	\end{figure}
	
	%We now address the $\union$ case. To aid the intuition behind these constructions, notice that the \dsabbr structure with the $\eps$-node and the empty-node define sets in $2^{\infAlph^*}$, and can be thought of as the semiring $(2^{\infAlph^*}\!\!, \emptyset, \{\eps\}, \times, \cup)$. In particular 
	
	Similarly, we address the operation $\union(v_3, v_4) \to v'$ by considering each case separately.
	%Since this is a commutative operation, in the sense that the result of $\union( v_3, v_4)$ and $\union( v_4, v_3)$ should define the same set, we only describe the cases where the conditions satisfied by $v_3$ and $v_4$ are $(1,1)$, $(1, 2)$, $(1, 3)$, $(2, 2)$, $(2, 3)$ and $(3, 3)$.
	\begin{itemize}
		\item Assume $v_3$ is the $\eps$-node (condition 1). (1) If $v_4$ also is the $\eps$-node, then we simply return $v'\gets\eps$. (2) If $v_3$ satisfies condition 2, let $v'$ be a union node such that $\ell(v') = \eps$ and $r(v') = v_4$. It can be seen that $v'$ satisfies condition 3 and that its $\eps$-safe. (3) If $v_4$ satisfies condition 3, we simply return $v'\gets v_4$.
		\item Assume $v_3$ satisfies condition 2. (1) If $v_4$ is the $\eps$-node, we return $v'\gets \union( v_4, v_3)$ using the construction for the case (2) in the previous item. (2) If $v_4$ satisfies condition 2, then we return $v'\gets \union( v_3, v_4)$ as it was defined for \dsabbrs without $\eps$. (3) If $v_4$ satisfies condition 3, let $v'' \gets \union( v_3, r(v_4))$, and let $v'$ be a union node such that $\ell(v') = \eps$ and $r(v) = v''$. Again, it can be seen that $v'$ is $\eps$-safe.
		\item Assume $v_3$ satisfies condition 3. (1) If $v_4$ is the $\eps$-node, we simply return $v'\gets\eps$. (2) If $v_4$ satisfies condition 2, we follow the construction given in case (3) of the previous item. (3) If $v_4$ satisfies condition 3, let $v''\gets\union( r(v_3), r(v_4))$ and let $v'$ be a union node such that $\ell(v') = \eps$ and $r(v) = v''$. It can be seen that $v'$ is $\eps$-safe.
	\end{itemize}
	
	Clearly, in each of these cases above we have that $\sem{\cD}(v') = \sem{\cD}(v_3)\cup \sem{\cD}(v_4)$, and if $v_3$ and $v_4$ are $\eps$-safe, then $v'$ is $\eps$-safe as well. In addition, they all take constant time and in each case the new version of $\cD$ satisfies the $\eps$-condition.
	
	The operation $\shiftop(v, s)$ is defined as follows. If $v$ is the $\eps$-node, we return $v'\gets\eps$. If $v$ satisfies condition 2, we return $v'\gets\shiftop(v, s)$ as it was defined for \dsabbrs without $\eps$. If $v$ satisfies condition 3, let $v''\gets\shiftop(r(v), s)$ and let $v'$ be a union node such that $\ell(v') =\eps$ and $r(v) = v''$. It can be seen that $v'$ is $\eps$-safe.
	
	By the previous constructions, we can conclude that the operations $\add$, $\prod$, $\union$ and $\shiftop$, when done over $\eps$-safe nodes, can be done in constant time, return an $\eps$-safe node, maintain the $\eps$-condition in the \dsabbr, and satisfy the specifications of each operation. Furthermore, by virtue of Corollary~\ref{prop:lindelayeps}, the set $\sem{D}(v)$ can be enumerated with output-linear~delay for each node $v$ in $\cD$.

%We summarize the case of $\epsilon$-nodes in the following theorem which is an extension of Theorem~\ref{theo:data-structure}.

\begin{thm}\label{theo:data-structure-eps}
	The operations $\add$, $\prod$, $\union$ and $\shiftop$ over \dsabbr{} extended with empty- and $\epsilon$-nodes take constant time. Furthermore, if we start from an empty \dsabbr{}~$\cD$ and apply $\add$, $\prod$, $\union$, and $\shiftop$ over $\epsilon$-safe nodes, the resulting node $v'$ is always an $\epsilon$-safe node, and the set $\sem{\cD}(v)$ can be enumerated with output-linear delay without preprocessing for every node $v$.
\end{thm} 

For the rest of the paper, we assume that a \dsabbr{} is a tuple $\cD = (\infAlph, V, \lch, \rch, \lambda, \bot, \epsilon)$ where we define $\infAlph, V, \lch, \rch,\lambda$ as before, and $\bot, \epsilon \in V$ are the empty- and $\epsilon$-node, respectively. Further, we assume that $\lch$, $\rch$, and $\lambda$ are extended accordingly, namely,  $\lch(v)$ and $\rch(v)$ are not defined whenever $v \in \{\bot, \epsilon\}$, and $\lambda:V\to\infAlph\cup\int \cup \{\cup, \odot,\bot, \eps\}$ such that $\lambda(v) = \bot$ ($\lambda(v) = \epsilon$) iff $v = \bot$ ($v = \eps$, respectively). 
%Finally, we can extend Theorem~\ref{theo:data-structure} for the \dsabbr{} extension as follows. 

	\section{Evaluation of annotated automata over SLP-compressed strings}\label{sec:evaluation}
	
	%!TEX root = ../main/main.tex

This section shows our algorithm for evaluating an annotated automaton over an SLP-compressed document. This evaluation is heavily inspired by the preprocessing phase in~\cite{SchmidS21}, as it primarily adapts the algorithm to the \dsabbr{} data structure. In a nutshell, we keep matrices of \dsabbr{} nodes, where each matrix represents the outputs of all partial runs of the annotated automaton over fragments of the compressed strings. We extend the operations of \dsabbr{} over matrices of nodes, which will allow us to compose matrices, and thus compute sequences of compressed strings. Then the algorithm proceeds in a dynamic programming fashion, where matrices are computed bottom-up for each non-terminal symbol. Finally, the matrix that corresponds to the start symbol of the SLP will contain all the outputs. The result of this process is that each matrix entry succinctly represents an output set 
%can be operated in constant time, 
that can enumerated with output-linear delay.

%For the rest of this section, we start by extending \dsabbr{} for matrices of nodes to finish with the presentation of the main algorithm. 

\paragraph{Matrices of nodes} The main ingredient for the evaluation algorithm are matrices of nodes for encoding partial runs of annotated automata. To formalize this notion, fix an unambiguous \rt $\cA = (Q, \Sigma, \oalph, \Delta, \qinit, F)$ and a \dsabbr{} $\cD = (\infAlph, V, \lch, \rch, \lambda, \bot, \epsilon)$. We define a \textdef{partial run} $\rho$ of $\cA$ over a document $d = a_1a_2\ldots a_n \in\Sigma^*$ as a sequence 
\[
\rho \ := \ p_1 \xrightarrow{b_1} \ldots \xrightarrow{b_n} p_{n+1}
\]
such that $p_1\in Q$, and for each $i\in[1,n]$ either $b_i = a_i$ and $(q_{i}, a_i, q_{i+1})\in \Delta$, or $b_i = (a_i, \oout)$ and $(q_{i}, (a_i, \oout), q_{i+1})\in \Delta$. Additionally, we say that the partial run $\rho$ is from state $p$ to state $q$ if $p_1 = p$ and $p_{n+1} = q$. In other words, partial runs are almost equal to runs, except they can start and end at any state $p$ and $q$, respectively.
 
For the algorithm, we use the set of all $Q\times Q$ matrices where entry $M[p,q]$ is a node in~$V$ for every $p,q \in Q$.
Each node $M[p,q]$ represents all annotations %$\sem{\cD}(M[p,q])$ 
of partial runs from state $p$ to  state $q$, which can be enumerated with output-linear delay by Theorem~\ref{theo:data-structure-eps}. Further, $M[p,q] = \bot$ represents that there is no run, and $M[p,q] = \epsilon$ that there is a single run without outputs (i.e., a run that produces the $\epsilon$ output). 

To combine these matrices of nodes, we define two operations.
The first operation is the \emph{matrix multiplication} over the semiring $(2^{(\oalph\times \int)^*}, \cup, \cdot, \emptyset, \{\epsilon\})$ but represented over $\cD$. Formally, let $Q = \{q_1, \ldots, q_{m}\}$ with $m = |Q|$. Then, for two $m \times m$ matrices $M_1$ and $M_2$, we define $M_1 \otimes M_2$ such that for every $p,q \in Q$:
\[
(M_1 \otimes M_2)[p, q] := \union_{i=1}^{m} \bigg( \prod\big(M_1[p, q_i], M_2[q_i, q]\big) \bigg)
\]
where $\union_{i=1}^{m} E_i := \union(\ldots \union(\union(E_1, E_2), E_3)\ldots,E_m)$. 
That is, the node $(M_1 \otimes M_2)[p, q]$ represents the set $\bigcup_{i=1}^{m} \big( \sem{\cD}(M_1[p, q_i]) \cdot \sem{\cD}(M_2[q_i, q]) \big)$.% which is constructed with operations over $\cD$.

The second operation for matrices is the extension of the \emph{shift operation}. Formally, $\shiftop(M,s)[p,q] := \shiftop(M[p,q], s)$ for a matrix $M$, $s \in \int$, and $p,q \in Q$.  
Since each operation over $\cD$ takes constant time, overall multiplying $M_1$ with $M_2$ takes time $\cO(|Q|^3)$ and shifting $M$ by $s$ takes time $\cO(|Q|^2)$.

\let\oldReturn\Return
\renewcommand{\Return}{\State\oldReturn}
\newcommand{\lenx}{\operatorname{len}}

\begin{algorithm*}[t]
	\caption{The enumeration algorithm of an unambiguous AnnA $\cA = (Q,\Sigma, \oalph, \Delta,\qinit,F)$ over an SLP $S = (N, \Sigma, R, S_0)$.}\label{alg:evaluation}
	\smallskip
	\begin{varwidth}[t]{0.60\textwidth}
		\begin{algorithmic}[1]
			\Procedure{Evaluation}{$\cA,S$}
			\State Initialize $\cD$ as an empty \dsabbr
			\State $\textproc{NonTerminal}(S_0)$
			\State $v \gets \bot$
			\ForEach{$p\in\qinit, q \in F$} 
			\State $v \gets \union(v, M_{S_0}[p, q])$
			\EndFor
			\State $\textproc{Enumerate}(v, \cD)$
			
			\EndProcedure
			\smallskip
			
			\Procedure{Terminal}{$a$}
			\State $M_a \gets \{[p,q] \to \bot \mid p,q \in  Q\}$
			\ForEach{$(p,(a,\oout), q) \in \Delta$} 
			\State $M_a[p,q] \gets \union(M_a[p,q],\, \add(\oout))$
			\EndFor
			\ForEach{$(p,a,q) \in \Delta$} 
			\State $M_a[p,q] \gets \union(M_a[p,q], \, \epsilon)$
			\EndFor
			\State $\lenx_a \gets 1$
			\EndProcedure
			\algstore{myalg}
		\end{algorithmic}
	\end{varwidth} \ \ \ \ 
	\begin{varwidth}[t]{0.60\textwidth}
		\begin{algorithmic}[1]
			\algrestore{myalg}
			\Procedure{NonTerminal}{$X$}
			\State $M_X \gets \{[p,q] \to \bot \mid p,q \in  Q, p \neq q\} \, \cup$ \par\hspace{2.7em} $\{[p,q] \to \epsilon \mid p,q \in  Q, p = q\}$
			\State $\lenx_X \gets 0$
			\For{$i=1$ \textbf{to} $|R(X)|$}
			\State $Y \gets R(X)[i]$
			\If{$M_Y$ is not defined}
			\If{$Y \in \Sigma$}
			\State $\textproc{Terminal}(Y)$
			\Else
			\State $\textproc{NonTerminal}(Y)$
			\EndIf 
			\EndIf 
			\State $M_X \gets M_X \otimes \shiftop(M_Y, \lenx_X)$
			\State $\lenx_X \gets \lenx_X + \lenx_Y$
			\EndFor 
			\EndProcedure
		\end{algorithmic}
	\end{varwidth} 
	\smallskip
\end{algorithm*}

\paragraph{The algorithm} We present the evaluation algorithm for the $\enumraa$ problem in Algorithm~\ref{alg:evaluation}. As expected, the main procedure \textproc{Evaluation} receives as input an unambiguous annotated automaton $\cA = (Q,\Sigma, \oalph, \Delta,\qinit,F)$ and an SLP  $S = (N, \Sigma, R, S_0)$, and enumerates all outputs in $\sem{\cA}(\doc(S))$. To simplify the notation, in Algorithm~\ref{alg:evaluation} we assume that $\cA$ and $S$ are globally defined, and we can access them in any subprocedure. Similarly, we use a \dsabbr{} $\cD$, and, for every $X \in N \cup \Sigma$, a matrix $M_X$ and an integer $\lenx_X$, which can globally be accessed at any place as well.

The main purpose of the algorithm is to compute $M_X$ and $\lenx_X$ recursively. On one hand, $M_X$ is a $Q\times Q$ matrix where each node entry $M_X[p,q]$ represents all outputs of partial runs from $p$ to $q$. On the other hand, $\lenx_X$ is the length of the string $R^*(X)$ (i.e., the string produced from $X$). Both $M_X$ and $\lenx_X$ start undefined, and we compute them recursively, beginning from the non-terminal symbol $S_0$ and by calling the method $\textproc{NonTerminal}(S_0)$~(line 3). After $M_{S_0}$ was computed, we can retrieve the set $\sem{\cA}(S)$ by taking the union of all partial run's outputs from an initial state $p\in\qinit$ to a state $q \in F$, and storing it in node $v$ (lines 4-6). Finally, we can enumerate  $\sem{\cA}(S)$ by enumerating all outputs represented by~$v$~(line 7). The reason why these outputs can be printed without repetitions, in short, is because every (relevant) call to {\sf union} and {\sf prod} is done while respecting the conditions of a duplicate-free \dsabbr. The details are in the proof of Theorem~\ref{theo:evaluation}, in a dedicated paragraph titled {\em duplicate-freeness}.

The workhorses of the evaluation algorithm are procedures \textproc{NonTerminal} and \textproc{Terminal} in Algorithm~\ref{alg:evaluation}. The former computes matrices $M_X$ recursively whereas the latter is in charge of the base case $M_a$ for a terminal $a \in \Sigma$. For computing the base case, we can start with $M_a$ with all entries equal to the empty node $\bot$ (line 9). Then if there exists a read-annotate transition $(p, (a, \oout), q) \in \Delta$, we add an output node $\oout$ to $M_a[p,q]$, by making the union between the current node at $M_a[p,q]$ with the node $\add(\oout)$ (line 11). Also, if a read transition $(p, a, q) \in \Delta$ exists, we do the same but with the $\epsilon$-node (line 13). Finally, we set the length of $\lenx_a$ to $1$, and we have covered the base case. 

For the recursive case (i.e., procedure $\textproc{NonTerminal}(X)$), we start with a sort of ``identity matrix'' $M_X$ where all entries are set up to the empty-node except the ones where $p = q$ that are set up to the $\eps$-node, and the value $\lenx_X = 0$ (lines 16-17). Then we iterate sequentially over each symbol $Y$ of $R(X)$, where we use $R(X)[i]$ to denote the $i$-th symbol of $R(X)$ (lines 18-19).
If $M_Y$ is not defined, then we recursively compute $\textproc{Terminal}(Y)$ or $\textproc{NonTerminal}(Y)$ depending on whether $Y$ is in $\Sigma$ or not, respectively (lines 20-24). The matrix $M_Y$ is memorized (by having the check in line 20 to see if it is defined or not) so we need to compute it at most once.
After we have retrieved $M_Y$, we can compute all outputs for $R(X)[1] \ldots R(X)[i]$ by multiplying the current version of $M_X$ (i.e., the outputs of $R(X)[1] \ldots R(X)[i-1]$), with the matrix $M_Y$ shifted by the current length $\lenx_X$ (line~25).  Finally, we update the current length of $X$ by adding $\lenx_Y$ (line 26).

In the next theorem, we prove the correctness of~Algorithm~\ref{alg:evaluation}.

\begin{thm}\label{theo:evaluation}
	Algorithm~\ref{alg:evaluation} enumerates the set $\sem{\cA}(S)$ correctly for every unambiguous \rt $\cA$ and every SLP-compressed document $S$, with output-linear delay and after a preprocessing phase that takes time $\cO(|\cA| + |S|\times|Q|^3)$.
\end{thm}
\begin{proof}
%	Regarding correctness, the algorithm follows a direct matrix evaluation over the SLP grammar, where its correctness depends on the \dsabbr{} $\cD$. Notice that, although all operations over nodes are not necessarily duplicate-free, we know that the runs from an initial state $q\in\qinit$ to the final states are unambiguous. As a consequence, the operations used for the final output are duplicate-free.
	
	To aid in organization, we separate this proof in three main parts. First, in {\it correctness}, we prove that the node $v$ called in line 7 satisfies that $\sem{\cD}(v) = \sem{\cA}(S)$. Second, in {\it duplicate-freeness} we prove why the algorithm enumerates without repetitions. Third, in {\it time bounds} we prove that the preprocessing phase has the stated time bounds.
	
	{\it Correctness.}
	Our goal here is to prove that for each $M_X$, for an $X\in\Sigma\cup N$ that is reachable from $S_0$, the set $\sem{\cD}(M_X[p,q])$ contains the annotations of all partial runs of $\cA$ over $\doc(X)$ that start on $p$ and end on $q$
	
	First, let $X = a\in\Sigma$. This case can be quickly verified by inspecting the procedure {\sc Terminal}$(a)$: for each $p$ and $q$, the set $\sem{\cD}(M_a[p,q])$ contains what we expect, which is $\eps$ when $(p,a,q)\in\Delta$, and $(\oout, 1)$ for each $(p,(a,\oout),q)\in\Delta$. 
	
	Now, let $X$ be a non-terminal with $R(X) = Y_1\ldots Y_m$.
	Our goal is to prove by induction that on iteration $i$ on line 18, the set $\sem{\cD}(M_X[p,q])$ is made of the partial annotations of all runs over $\doc(Y_1\ldots Y_i)$ from $p$ to $q$.
	The base case here is actually when $i = 0$, and we consider lines 16-17 as the 0-th iteration and $\eps$ as the prefix of size 0 of $R(X)$. Thus, the sets $\sem{\cD}(M_X[p,q])$ on the 0-th iteration are what we expect, which is $\{\eps\}$ when $p = q$ and $\emptyset$ when $p\neq q$.
	Now, assume the statement holds for $i$. From the global inductive assumption, we have that the matrix $M_{Y_{i+1}}$ is correctly built, that is, that each set $\sem{\cD}(M_{Y_{i+1}}[p,q])$ contains the partial annotations of all runs over $\doc(Y_{i+1})$ from $p$ to $q$.
	By seeing that the operator $\otimes$ in line 25 indeed generates the matrix that corresponds to the join of $\sem{\cD}(M_X[p,q'])$ and $\shift(\sem{\cD}(M_{Y_{i+1}}[q',q]),|Y_1\ldots Y_i|)$ for each state $q'$, we conclude that the resulting matrix $M_X$ at the end of the iteration is correct.  
	From this reasoning, we conclude that the set $\sem{\cD}(M_X[p,q])$ contains exactly the annotations stated above.
	
	{\em Duplicate-freeness.} 
	 Here, we would hope to prove that the operators $\add$, $\union$, $\prod$ and $\shiftop$ are only called when the properties of a duplicate-free \dsabbr are true (see Section~\ref{sec:ecs-op}). However, this is not necessarily true, at least for the $\union$ case: $\cA$ is allowed to be ambiguous in partial runs that do not form part of an accepting run. For simplicity, the rest of the proof is stated only for indices $M_X[p,q]$ which are reached when building an output string. In other words, those where there exists an accepting run $\rho = p_1 \xrightarrow{b_1} \ldots \xrightarrow{b_n} p_{n+1}$ such that $p = p_\ell$, $q = p_r$ with $\ell \leq r$, and  $a_1\ldots a_{\ell-1}Xa_r\ldots a_n$ is reachable from $S_0$ by following production rules in $R$.
	
	In the case of $\prod$, proving the duplicate-free property is straightforward: it is only done in line 25, and the positions on the left side (i.e., the ones taken from $M_X$) are all less or equal than the value $\lenx_X$ has at that point. Thus, by shifting the values from $M_Y$ by this amount, the sets from either side will not contain any annotation in common. In the case of $\union$, one can check by inspection that in lines 11 and 13, this is also done properly: none of the transitions appears more than once, and one can check that in line 25, if any of the calls to $\union$ done inside the $\otimes$ operation is done with duplicates, then there is some annotation that appears in two distinct calls. This is because each of the $\union$ calls is done over a different index in the matrices $M_Y$ and $M_X$,  contradicting the assumption that $\cA$ is unambiguous. Therefore, we conclude that the operators $\add$, $\union$, $\prod$ and $\shiftop$ are called properly, and the statement of the theorem follows.
	
	{\it Time bounds.} The main procedure calls \textproc{NonTerminal} or \textproc{Terminal} at most once for every symbol. After making all calls to \textproc{Terminal}, each transition in $\Delta$ is seen exactly once, and \textproc{NonTerminal} takes time at most $\cO(|R(X)|\times|Q|^3)$ not taking into account the calls inside. Adding these values gives us an overall $\cO(|\cA| + |S|\times|Q|^3)$ execution time.
\end{proof}

We want to finish by noticing that, contrary to~\cite{SchmidS21}, our evaluation algorithm does not need to modify the grammar $S$ into Chomsky's normal form (CNF) since we can evaluate $\cA$ over $S$ directly. %Although passing $S$ into CNF can be done in linear time over $S$~\cite{SchmidS21}, this step can incur an extra cost, which we can avoid in our approach. 

	\section{Applications in regular spanners}\label{sec:spanners}
	
	%!TEX root = ../main.tex

It was already shown in~\cite{AmarilliJMR22} that working with annotations directly and then providing a reduction from a spanner query to an annotation query is sometimes more manageable. In this section we will do just that: starting from a document-regular spanner pair $(d, \mathcal{M})$, we will show how to build a document-annotated automaton pair $(d', \cA)$ such that $\mathcal{M}(d) = \sem{\cA}(d')$. Although people have studied various models of regular spanners in the literature, we will focus here on sequential variable-set automata (VA)~\cite{FaginKRV15} and sequential extended variable-set automata (eVA)~\cite{FlorenzanoRUVV20}. The latter appears to be syntactically equivalent to the model that Schmid and Schweikardt used in their results ~\cite{SchmidS21}. We present the models in this order, but we present first the reduction from the latter as it can be done into \rt directly.
In the second half of the section we reduce the former to {\it succinctly} annotated automata, an extension of \rt that allows output symbols to be stored concisely. 
These reductions imply output-linear delay enumeration for the spanner~tasks.

Let it be noted that, when dealing with spanners, \emph{output-linear delay} is indeed equivalent to what is commonly known as \emph{constant delay}. 
We believe the latter to be somewhat ambiguous, even in these tasks: the size of an output depends on the number of variables, and this is given by the query.
Therefore, the term is accurate only when looking at them in data complexity---not, say, in combined or query complexity.
We choose to continue talking in terms of \emph{output-linear delay} for the rest of the section to resolve any imprecisions, while reminding the reader that the results are fully compatible with those called \emph{constant delay} in the literature.
%For the former we show a reduction that achieves output-linear delay in the second half of this section. For this result, we will need to introduce an extension of annotated automata that allows output symbols to be represented succinctly.

\paragraph{Variable-set automata} Consider a document $d = a_1\ldots a_n$ over an input alphabet $\Sigma$. A {\it span} of $d$ is a pair $\spanc{i}{j}$ with $1 \leq i \leq j \leq n+1$. We define the substring at $\spanc{i}{j}$ by $d\spanc{i}{j} = a_i \ldots a_{j-1}$. We also consider a finite set of variables $\cX$ and we define a \emph{mapping} as a partial function that maps some of these variables to spans. We define a \emph{document spanner} as a function assigning every input document $d$ to a set of mappings~\cite{FaginKRV15}.

A \emph	{variable-set automaton} (VA for short) is a tuple $\cA = (Q, \Sigma, \cX, \Delta, \qinit, F)$ where $Q$ is a set of states, $\qinit,F\subseteq Q$, and $\Delta$ consists of {\it read transitions} $(p,a,q)\in Q\times\Sigma\times Q$ and {\it variable transitions} $(p, \varop{x}, q)$ or $(p,\varcl{x},q)$ where $p, q\in Q$ and $x\in \cX$.
The symbols $\varop{x}$ and $\varcl{x}$ are referred to as {\it variable markers} of $x$, where $\varop{x}$ is {\it opening} and $\varcl{x}$ is {\it closing}. Given a document $d = a_1\ldots a_n \in \Sigma^*$ a configuration of $\cA$ is a pair $(q,i)$ where $q\in Q$ and $i\in [1, n+1]$. 
A run $\rho$ of $\cA$ over $d$ is a sequence:
\[
\rho \ := \ (q_1, i_1)\xrightarrow{\sigma_1} (q_2,i_2) \xrightarrow{\sigma_2}\cdots\xrightarrow{\sigma_m}(q_{m+1}, i_{m+1})
\] 
where $i_1 = 1$, $i_{m+1} = n+1$, and for each $j\in[1,m]$, $(q_j, \sigma_{j},q_{j+1}) \in \Delta$ and either 
(1) $\sigma_{j} = a_{i_j}$ and $i_{j+1} = i_j + 1$, or (2) $\sigma_{j}\in \{\varop{x} , \varcl{x}\mid x\in\cX\}$ and $i_{j+1} = i_j$. We say that $\rho$ is {\it accepting} if $q_{m+1}\in F$ and that it is {\it valid} if variables are non-repeating, and they are opened and closed correctly. 
%Namely, $\rho$ is valid if for each $x\in\cX$, $\varop{x}$ appears at most once, and if it does, $\varcl{x}$ appears someplace to the right. 
If $\rho$ is accepting and valid, we define the mapping $\mu^\rho$ which maps $x\in \cX$ to the span $\spanc{u}{v}$ if, and only if, there exist $j,k\in[1,m]$ such that $i_j = u, i_k = v$, and $\sigma_j =\, \varop{x}$ and $\sigma_k = \varcl{x}$. We say that $\cA$ is {\it sequential} if every accepting run is also valid. Finally, define the document spanner $\sem{\cA}$ as the function:
\[
\sem{\cA}(d) \ = \  \{\mu^\rho\mid \rho \text{ is an accepting and valid run of $\cA$ over $d$}\}.
\] 
Like in \rts, we say $\cA$ is {\it unambiguous} if for each mapping $\mu\in\sem{\cA}(d)$ there is exactly one accepting run $\rho$ of $\cA$ over $d$ such that $\mu^\rho = \mu$.

\paragraph{Extended VA}
An \emph{extended variable-set automaton} (or eVA for short) is a tuple $\cA = (Q, \Sigma, \cX, \Delta,\qinit, F)$ where $Q,I,F$ are defined as in VA, $\Delta$ is a set consisting of {\it letter transitions} $(p, a, q)$ where $a\in\Sigma$ and $p,q\in Q$ or {\it extended variable transitions} $(p, S, q)$ where $S\subseteq \{\varop{x} , \varcl{x}\mid x\in\cX\}$ and $S$ is non-empty; and $F\subseteq Q$. A run $\rho$ over a document $d = a_1\ldots a_n$ is a sequence:
\[
\rho = q_1  \xrightarrow{S_1} p_1 \xrightarrow{a_1} 
q_2  \xrightarrow{S_2} p_2 \xrightarrow{a_2} \cdots 
\xrightarrow{a_n} q_{n+1}  \xrightarrow{S_{n+1}} p_{n+1}
\]
where each $S_i$ is a (possibly empty) set of markers, for each $i\in[1,n]$, $(p_i, a_{i}, q_{i+1})\in \Delta$, and for each $i\in[1,n+1]$, if $S_i$ is not empty, then $(q_i, S_{i}, p_i)\in \Delta$, and $p_i = q_i$, otherwise. We say that a run is accepting if $p_{n+1}\in F$. Also, we say that a run $\rho$ is {\it valid} if variables are opened and closed in a correct manner: every marker $\varop{x}$ and $\varcl{x}$ must appear at most once among the sets $S_1\ldots S_{n+1}$; if one of them appears, the other does as well; and if $\varop{x}\in S_i$ and $\varcl{x}\in S_j$ then it holds that $i\leq j$. For a valid run $\rho$, we define the mapping $\mu^\rho$ that maps $x$ to $\spanc{i}{j}$ iff $\varop{x}\in S_i$ and $\varcl{x}\in S_j$. The spanner $\sem{\cA}$ is defined identically as for 
VA. The definitions of sequential and unambigous eVA are the same as well.

To motivate the reduction from sequential eVA to annotated automata, consider a document $d = \texttt{aab}$, and a run over $d$ of some (unspecified) eVA with variable set $\cX = \{x, y\}$:
\[
\rho = q_1  \xrightarrow{\emptyset} q_1 \xrightarrow{\texttt{a}} 
q_2  \xrightarrow{\{\varop{x}, \varcl{x}, \varop{y}\}} p_2 \xrightarrow{\texttt{a}} 
q_3  \xrightarrow{\emptyset} q_3 \xrightarrow{\texttt{b}} 
q_4  \xrightarrow{\{\varcl{y}\}} p_4
\]
This run defines the mapping $\mu$ which assigns $\mu(x) = \spanc{2}{2}$ and $\mu(y) = \spanc{2}{4}$.
To translate this run to the annotated automata model, first we append an end-of-document character to~$d$ so we are then able to ``push'' the marker sets one transition to the right. That way, we obtain a possible run of an annotated automaton with output set $\oalph = 2^{\{\varop{x} , \varcl{x}\mid x\in\cX\}}$ over the document $d' = \texttt{aab\#}$:
\[
\rho' = q_1'  \xrightarrow{\texttt{a}} 
q_2'  \xrightarrow{(\texttt{a}, \{\varop{x}, \varcl{x}, \varop{y}\})} 
q_3'  \xrightarrow{\texttt{b}} 
q_4'  \xrightarrow{(\texttt{\#}, \{\varcl{y}\})} q_5'
\]
The annotation of this run would then be $(2, \{\varop{x}, \varcl{x}, \varop{y}\})(4,\{\varcl{y}\})$, from where the mapping $\mu$ can be extracted directly. The reduction from eVA into annotated automata operates in a similar fashion: the read transitions are kept, and for each pair of transitions $(p,S,q), (q, a, r)$ in the former, a transition $(p, (a, S), r)$ is added to the latter.

%The idea for the reduction from extended VA to annotated automata can be seen more clearly with help of subword-marked words, introduced in~\cite{SchmidS21}. Let us consider the document $d = \texttt{aabba}$. Given a set of markers $\cX = \{x, y\}$, a subword-marked word for $d$ would be $\bar{d} = \texttt{a}\texttt{a}\{\varop{x}\}\texttt{b}\texttt{b}\{\varcl{x}, \varop{y}, \varcl{y}\}\texttt{a}$, from which we can extract the mapping $\mu$ which assigns $\mu(x) = \spanc{3}{5}$ and $\mu(y) = \spanc{5}{5}$. These strings can also be seen as a possible run from an extended VA over $d$, where the states have been shaved off and only what is read on the transitions is kept.
%Now let us add an end-of-document character $\texttt{\#}$ to $\bar{d}$ and replace every appearance of $Sa$, for some set of markers $S$ and $a\in \Sigma$, by the pair $(a, S)$. This renders the string $\hat{d} = \texttt{a}\texttt{a}(\texttt{b}, \{\varop{x}\})\texttt{b}(\texttt{a}, \{\varcl{x}, \varop{y}, \varcl{y}\})\texttt{\#}$. Note that this word can now be seen as a possible run from an annotated automaton over $d$, if its set of outputs has been defined as $\oalph = 2^{\{\varop{x} , \varcl{x}\mid x\in\cX\}}$. The annotation of this run would then be $\nu = (3, \{\varop{x}\})(5, \{\varcl{x}, \varop{y}, \varcl{y}\})$, from where the mapping $\mu$ can be extracted directly. This idea of ``pushing'' the marker sets to the next letter can be applied in the automaton itself, which is what the reduction does.

The equivalence between mappings and annotations is formally defined as follows: for some document $d$, a mapping $\mu$ from $\cX$ to spans in $d$ is equivalent to an annotation $w = (S_1,i_1)\ldots(S_m, i_m)$ if, and only if, for every $j\in[1,m]$: 
\[
S_j \ = \ \{\,\varop{x} \,\mid \mu(x) = \spanc{i_j}{i'} \text{ for some } i'\} \, \cup \, \{\,\varcl{x} \,\mid \mu(x) = \spanc{i'}{i_j}\text{ for some } i'\}.
\]
\begin{prop}\label{prop:eva-reduc}
	For any unambiguous sequential eVA $\cA$ with state set $Q$ and transition set $\Delta$, there exists an \rt $\cA'$ with $\cO(|Q|\times|\Delta|)$ transitions 	such that each mapping $\mu\in \sem{A}(d)$ is equivalent to some unique $w\in\sem{\cA'}(d\texttt{\#})$ and vice versa, for every document~$d$.
\end{prop}
\begin{proof}
	Let $\cA = (Q, \Sigma, \cX, \Delta, \qinit, F)$ be an unambiguous sequential eVA. We build an annotated automaton $\cA' = (Q', \Sigma\cup\{\texttt{\#}\}, \oalph, \Delta', \qinit, F')$ as follows.  Define $Q' = Q \cup \{q^*\}$ for some fresh state~$q^*$, and $F' = \{q^*\}$. Further, define $\oalph = 2^{\{\varop{x} , \varcl{x}\mid x\in\cX\}}$ and: 
	\begin{align*}
		\Delta' = \ &\{(p,a,q)\mid a\in \Sigma\text{ and }(p,a,q)\in\Delta\}\  \cup\\ & \{(p,(a, S), q)\mid (p,S,q'),(q',a,q)\in\Delta \text{ for some }q'\in Q\}\ \cup\\ & \{(p,\texttt{\#}, q^*)\mid p\in F\}\ \cup\\ & \{(p,(\texttt{\#}, S), q^*)\mid (p,S,q)\in\Delta\text{ for some }q\in F\}.
	\end{align*}
	To see the equivalence between $\cA$ and $\cA'$, let $d = a_1\ldots a_n$ be a document over $\Sigma$, and let $\rho$ be an accepting run of $\cA$ over $d$ of the form:
	\[
	\rho = q_1 \xrightarrow{S_1} p_1 \xrightarrow{a_1} 
	q_2  \xrightarrow{S_2} p_2 \xrightarrow{a_2} \cdots 
	\xrightarrow{a_n} q_{n+1} \xrightarrow{S_{n+1}} p_{n+1}
	\]
	We define $\rho'$ as the following sequence:
	\[
	\rho' = q_1  \xrightarrow{b_1} 
	q_2  \xrightarrow{b_2} \cdots 
	\xrightarrow{b_n} q_{n+1}  \xrightarrow{b_{n+1}} q^*
	\]
	where $b_i  = (a_i, S_i)$ if $S_i$ is not empty, and $b_i = a_i$ otherwise, for each $i\leq n$. We define $b_{n+1} = (\texttt{\#}, S_{n+1})$ if $S_{n+1}$ is not empty, and $b_{n+1} = \texttt{\#}$ otherwise. Since $\cA$ is sequential, $\rho$ is a valid run which defines a mapping $\mu^\rho\in\sem{\cA}(d)$. We can straightforwardly check that $\mu^\rho$ is equivalent to $\ann(\rho')$.
	It can also be seen directly from the construction that $\rho'$ is a run from $\cA'$ over $d\texttt{\#}$, and since $\rho'$ is uniquely defined from $\rho$, we conclude that for every document $d$ each $\mu\in\sem{\cA}(d)$ is equivalent to some unique $w\in\sem{\cA'}(d\texttt{\#})$.
	
	To see the equivalence on the opposite direction, consider an accepting run $\rho'$ of $\cA'$ over $d\texttt{\#}$ as above, where each $b_i$, for $i\in[1, n]$, might be either $a_i$ or a pair $(a_i, S)$. 
	From the construction, it can be seen that if $b_i = a_i$, there exists a transition $(q_{i},a_i, q_{i+1})\in\Delta$. Instead, if $b_i = (a_i, S)$, there exist transitions $(q_{i},S, q'),(q',a_i,q_{i+1})\in\Delta$ for some $q'\in Q$. 
	Also, if $b_{n+1} = \texttt{\#}$ then $q_{n+1}\in F$, and if $b_{n+1} = (\texttt{\#}, S)$ there exists $(q_{n+1},S,q')\in\Delta$ for some $q'\in F$. We define $\rho$ as a run of $\cA$ over $d$ built by replacing each transition in $\rho'$ by the corresponding transition(s) in $\cA$. We note first that this run is accepting and valid, and since $\cA$ is unambiguous, $\rho$ must be uniquely defined. 
	Indeed, when replacing $(q_{i}, (a_i, S), q_{i+1})$, we know there exist transitions $(q_{i}, S, q')$ and $(q', a_i, q_{i+1})$ in $\Delta$. 
	Furthermore, this $q'$ must be unique, otherwise we could define a different accepting run that defines the same mapping. 
	We see that $\ann(\rho')$ is equivalent to $\mu^\rho$, so we conclude that each annotation $w\in\sem{\cA}(d)$ is equivalent to some unique mapping $\mu\in\sem{\cA}(d)$.
	
	To see that $\cA'$ is unambiguous, consider towards a contradiction two different accepting runs $\rho_1'$ and $\rho_2'$ that retrieves the same annotation. Let $i$ be such that the $i$-th states in $\rho_1'$ and $\rho_2'$ are different, and note it holds that $1\leq i \leq n+1$ since $\cA'$ has a unique final state $q^*$. 
	By the previous discussion, we can build runs $\rho_1$ and $\rho_2$ of $\cA$ over $d$ that also define the same mapping. Furthermore, they differ at index $2i-1$ (the index at which $q_i$ is in the $\rho$ written above), which is not possible since $\cA$ is unambiguous.
	\qedhere
\end{proof}

Combining Proposition~\ref{prop:eva-reduc} and Theorem~\ref{theo:main-theorem}, we get an output-linear delay algorithm for evaluating an unambiguous sequential eVA over a document, proving the extension of the result in~\cite{SchmidS21}.
\begin{cor}\label{coro:eva}
	Given an unambiguous sequential eVA $\cA$ and an SLP $S$, there exists an enumeration algorithm that runs in $|\cA|^3\times |S|$-preprocessing time and output-linear delay for enumerating $\sem{\cA}(\doc(S))$.
\end{cor}  

Notice that the result in~\cite{SchmidS21} is for \emph{deterministic} VA, where here we generalize this result for the \emph{unambiguous} case, plus the output-linear delay.

\paragraph{Succinctly annotated automata}
For the algorithmic result of sequential (non-extended) VA, we need an extension to annotated automata which features succinct representations of sets of output symbols. 

A {\it succinct enumerable representation scheme} (SERS) is a tuple: 
\[
\cS = (\cR, \infAlph, |\cdot|, \cL, \cE)
\]
made of an infinite set of representations $\cR$, and an infinite set of output elements $\infAlph$. 
It includes a function $|\cdot|$ that indicates, for each $r\in\cR$ and $\oout\in\infAlph$, the sizes $|r|$ and $|\oout|$, i.e., the number of units needed to store $r$ and $\oout$ in the underlying computational model (e.g., the RAM model).
The function $\cL$ maps each element $r\in\cR$ to some finite non-empty set $\cL(r)\subseteq\infAlph$. 
Lastly, there is an algorithm $\cE$ which enumerates the set $\cL(r)$ with output-linear delay for every $r\in \cR$. 
Intuitively, a SERS provides us with representations to encode sets of outputs. Moreover, there is the promise of the enumeration algorithm $\cE$ where we can recover all the outputs with output-linear delay. This representation scheme allows us to generalize the notion of annotated automaton for encoding an extensive set of sequences of output symbols in the~transitions.

Fix a SERS $\cS = (\cR, \infAlph, |\cdot|, \cL, \cE)$. A Succinctly Annotated Automaton over $\cS$ (\crt for short) is a tuple $\cA = (Q, \Sigma, \oalph, \Delta, \qinit, F)$ where all sets are defined like in \rt, except that in $\Delta$ read-annotate transitions are of the form $(p, (a, r), q)\in Q\times(\Sigma\times\cR)\times Q$. That is, transitions are now annotated by a representation $r$ that encodes a \emph{set} of elements in~$\infAlph$. For a read-annotate transition $\delta = (p, (a, r), q)$, we define its size as $|\delta| = |r| + 1$ and for a read transition $\delta = (p, a, q)$ we define its size as $|\delta| = 1$. 
A run $\rho$ over a document $d = a_1\ldots a_n$ is also defined as a sequence:
\[
\rho := q_1 \xrightarrow{b_1} q_2 \xrightarrow{b_2} \ldots \xrightarrow{b_n} q_{n+1}
\] 
with the same specifications as in \rt with the difference that it either holds that $b_i = a_i$, 
or $b_i = (a_i, r)$ for some representation $r$. We now define the {\it set of annotations of $\rho$} as: $\ann(\rho) = \ann(b_1, 1)\cdot\ldots\cdot\ann(b_n, n)$ where $ \ann(b_i, i) = \{(\oout, i)\mid\oout\in\cL(r)\}$ if $b_i = (a, r)$, and $\ann(b_i, i) = \{\eps\}$ otherwise.
%We use the concatenation product $\cdot$ over sets of strings in the natural way: $L_1\cdot L_2 = \{s_1\cdot s_2\mid s_1\in L_1 \text{ and }s_2\in L_2\}$. 
The set $\sem{\cA}(d)$ is defined as the union of sets $\ann(\rho)$ for all accepting runs $\rho$ of $\cA$ over $d$. We say that $\cA$ is unambiguous if for every document $d$ and every annotation $w\in\sem{\cA}(d)$ there exists only one accepting run $\rho$ of $\cA$ over $d$ such that $w\in\ann(\rho)$. Finally, we define the size of $\Delta$ as $|\Delta| = \sum_{\delta \in \Delta} |\delta|$, and the size of $\cA$ as $|\cA| = |Q| + |\Delta|$. 

Now we see how to use the fact that every set $\cL(r)$ can be enumerated with output-linear delay along with Theorem~\ref{theo:main-theorem} to evaluate a sAnnA over an SLP. This will result in the same bounds as before.

\begin{thm}\label{theo-algo-succ}
	Fix a SERS $\cS$. There exists an enumeration algorithm that, given an unambiguous \crt $\cA$ over $\cS$ with state set $Q$, and an SLP $S$, it runs in $(|\cA|+ |Q|^3 \times |S|)$-preprocessing time and output-linear delay for enumerating $\br{\cA}(\doc(S))$. 
\end{thm}
	We will work towards proving this theorem through the rest of this section.
	For this proof, we will use an extension of \dsabbrs we call {\it succinct \dsabbrs}.
	A succinct \dsabbrs is a tuple $\tilde{\cD} = (\cS, V, \lch, \rch, \lambda, \bot, \epsilon)$, similar to \dsabbr with the difference that the output set has been replaced by representations from a SERS $\cS$. The set of strings associated to a node $v\in \tilde{D}$ is defined as follows: 
	If $\lambda(v) = r\in \cR$, where $\cR$ is the set of representations in $\cS$, then $\sem{\tilde{D}}(v) = \cL(r)\times\{1\}$. The rest of the sets $\sem{\tilde{D}}(v)$ for union, product, and shift nodes $v$ remain unchanged, and so are the notions of duplicate-free, $k$-bounded and $\eps$-safe nodes, along with the operations for $\add$, $\prod$, $\union$ and $\shiftop$. 
	We will extend Proposition~\ref{theo:data-structure-eps} for this data structure as follows: 
	
	\begin{prop}\label{prop:ds-succ}
		The operations $\prod$, $\union$ and $\shiftop$ over succinct \dsabbrs{} extended with empty- and $\epsilon$-nodes take constant time, and the operation $\add$ with a representation $r$ as argument takes time $\cO(|r|)$. Furthermore, if we start from an empty succinct \dsabbr{} $\cD$ and apply $\add$, $\prod$, $\union$, and $\shiftop$ over $\epsilon$-safe nodes, the resulting node $v'$ is always an $\epsilon$-safe node, and the set $\sem{\cD}(v)$ can be enumerated with output-linear delay without preprocessing for every node $v$.
	\end{prop}
	\begin{proof}
		For this proof, we need to be a bit more specific regarding the enumeration algorithm $\cE$ of an SERS $\cS$. 
		Precisely, we assume that $\cE$ has an associated function $\yield$ and a constant $c$, and its procedure is to receive $r$ and then allow up to $|\cL(r)|$ calls to $\yield$. 
		Each of these calls produces a different $\oout\in \cL(r)$ and takes time at most $c\cdot (|\oout|+1)$ for some fix constant $c$. 
		Each of these output elements carries a flag $\texttt{end}$ which is true if, and only if, it is the last output of the set. 
		We also assume that a sequence of $|\cL(r)|$ calls to $\yield$ might happen again with the same time bounds after the last call had set $\texttt{end}$ to true.
		
		We start this proof by re-stating that every definition in the \dsabbr data structure is unchanged in its succinct version. The only difference is the definition of $\sem{\tilde{\cD}}(v)$ for a bottom node $v\in\tilde{\cD}$.
		
		First, we shall prove a version of Proposition~\ref{prop:lindelay} in this model, namely, that if a succinct \dsabbr $\tilde{\cD}$ is duplicate-free and $k$-bounded, then the set $\sem{\tilde{\cD}}(v)$ can be enumerated with output-linear delay for every $v\in \cD$. We do this by adapting the proof of Proposition~\ref{prop:lindelay} to handle this new type of bottom node. Let $\cS = (\cR, \infAlph, |\cdot |, \cL, \cE)$ be the SERS associated to $\tilde{\cD}$ and let $c'$ be the constant associated to $\cE$. 

		The main idea is to modify iterator $\tau_\Lambda$ from Algorithm~\ref{alg:enumeration} so that it stores, besides a value $u$ and a flag $\hasnext$, an output element $\oout$. The iterator fully makes use of the fact that the $\texttt{yield}$ procedure from $\cE$ retreives an output, evolves the internal iterator so that the next call produces the next output, and also says at each point if the current output was the last one from the set or not. The overall strategy is then to call $\texttt{yield}$ in $\scnext$, store the retrieved output in $\oout$, and in $\scprint$, print whatever is currently stored in $\oout$. More precisely, the $\sccreate$ procedure initializes $\cE$ so that the next time it calls $\texttt{yield}$ it prints the first output in the set $\cL(r)$, and initializes $\hasnext$ to {\bf true}. $\scnext$ first checks if $\hasnext$ is set to {\bf true}, and returns {\bf false} if it is not the case; otherwise it calls $\texttt{yield}$, stores the output in $\oout$, and returns {\bf true}. If the output of $\texttt{yield}$ is $\texttt{end}$, then it set $\hasnext$ to  {\bf false} and returns {\bf false}.   Finally, $\scprint$ with input $s$ simply prints the pair $(\oout, s+1)$. It can be seen that the methods follow the necessary specifications to ensure that the correctness of the algorithm still holds for this version of the data structure.
		
		To show the time bounds, we bring attention to the fact that the only difference in the algorithm time-wise is the time spent printing each pair $(\oout, i)$, as now this takes $c \cdot (|\oout| + 1)$. Since Algorithm~\ref{alg:enumeration} was proven to have delay $O(k\cdot(|w|+|w'|))$ to write output $w'$ after writing $w$ for non-succinct \dsabbr, we state that the same delay holds for this version of the algorithm since the time added is $O(|w|)$. This implies output-linear delay by following the same reasoning as the proof of Proposition~\ref{prop:lindelay}.
		
		The rest of the proof pertains the operations $\add$, $\prod$, $\union$ and $\shiftop$. It is not hard to see that they can be kept unchanged and maintain all the conditions in the statement, with the sole exception of $\add(r)$, which now adds a representation $r$ to the structure and takes $|r|$ time. This concludes the proof. 
	\end{proof}
	
	\begin{proof}[Proof of Theorem~\ref{theo-algo-succ}]
	The algorithm that we give to prove the statement is exactly Algorithm~\ref{alg:evaluation} line-by-line, with the sole exception of lines 10-11, which now read: 
	\[
	\begin{array}{l}
		\textbf{for each }(p,(a,r), q) \in \Delta \textbf{ do } \\
		\hspace{1cm} M_a[p,q] \gets \union(M_a[p,q],\, \add(r))
	\end{array}
	\] 
	The rest of the proof follows from the reasoning of the proof of Theorem~\ref{theo:evaluation}, and by noticing that the definition of the set $\sem{\tilde{D}}(v) = \cL(r) \times \{1\}$ satisfies what is expected for an index~$M_a[p,q]$. That is, that for every partial run $\rho$ of $\cA$ over $a$ that starts on $p$ and end on $q$, all annotations from the set $\ann(\rho)$ are included. The fact that the operations are done duplicate-free also follows from the fact that $\cA$ is assumed to be unambiguous.
\end{proof}

The purpose of \crt is to encode sequential VA succinctly. Indeed, as shown in~\cite{FlorenzanoRUVV20}, representing sequential VA by eVA has an exponential blow-up in the number of variables that cannot be avoided. Therefore, the reduction from Proposition~\ref{prop:eva-reduc} cannot work directly. Instead, we can use a Succinctly Annotated Automaton over some specific SERS to translate every sequential VA into the annotation realm efficiently. 

\begin{prop}\label{prop:va-reduc}
	There exists an SERS $\cS$ such that for any unambiguous sequential VA $\cA$ with state set $Q$ and transition set $\Delta$ there exists a \crt $\cA$ over $\cS$ of size $\cO(|Q|\times|\Delta|)$  such that for every document $d$, each mapping $\mu\in\sem{\cA}(d)$ is equivalent to some unique $w\in\sem{\cA}(d\texttt{\#})$ and vice versa. Furthermore, the number of states in $\cA$ is in $\cO(|Q|)$.
\end{prop}
\begin{proof}
	The SERS that we will consider are Enumerable Compact Sets as they were presented in~\cite{MunozR22}, defined over the set of output symbols $\{\varop{x} , \varcl{x}\mid x\in\cX\}$, with the sole difference that the sets of outputs stored in a node no longer contain strings, but subsets of $\{\varop{x} , \varcl{x}\mid x\in\cX\}$ instead. This difference is merely technical and has no influence in the time bounds of an ECS, as long as it is guaranteed to be duplicate-free (called {\it unambiguous} in~\cite{MunozR22}).
	 The set of output elements in our SERS will be $\infAlph = 2^{\{\varop{x} , \varcl{x}\mid x\in\cX\}}$, and the set of representations is given by the nodes in the ECS. 
	
	First we assume that all transitions in $\cA$ are reachable from some $q\in \qinit$, and all of them reach an accepting state. We know that in the VA $\cA$, the graph induced by the variable transitions is a DAG, otherwise it would not be sequential. We start with an empty ECS $\cD$ and define a matrix $K[p,q]$, that first starts with the empty node in each index except the indices in the diagonal (those that satisfy $p = q$), which have the $\eps$ node. 
	The idea is that at the end of this algorithm, $\cL_{\cD}(K[p,q])$ contains all sets of variable markers that can be seen in a path from $p$ to $q$ that does not contain a letter transition. We build this matrix by iterating over the variable transitions in $\cA$ following some topological order of the DAG, starting from a root. For each variable transition $(p,V,q)$, let $u\gets\add(V)$ and for each $p'\in Q$ that can reach $p$ in the DAG assign $K[p',q] \gets \union(K[p',q], \prod(K[p', p], u))$. The time of this algorithm is $|Q|\times|\cA|$.
	
	After doing this, we perform a construction analogous to the one done in Proposition~\ref{prop:eva-reduc}, in which we replace the extended variable transitions $(p,S,q)$ by transitions $(p,r,q)$ where $r = K[p,q]$, which happens every time $K[p,q]$ is not the empty node. The proof follows from this result as well.
\end{proof}

By Proposition~\ref{prop:va-reduc} and Theorem~\ref{theo-algo-succ} we prove the extension of the  output-linear delay algorithm for unambiguous sequential VA:
\begin{cor}\label{coro:va}
	Given an unambiguous sequential VA $\cA$ and an SLP $S$, there exists an enumeration algorithm that runs in $|\cA|^3\times |S|$-preprocessing time and output-linear delay for enumerating $\sem{\cA}(\doc(S))$.
\end{cor}  

	\section{Output-linear delay-preserving complex document editing}\label{sec:edits}
	
	%!TEX root = ../main.tex
% !TeX spellcheck = en_US

In this section, we show that the results obtained by Schmid and Schweikardt~\cite{SchmidS22} regarding enumeration over document databases and complex document editing still hold, maintaining the same time bounds in doing these edits, but allowing output-linear delay. We also include a refinement of the result for whenever the edits needed are limited to the concatenation of two documents. 

\paragraph{Complex document editing}
A document database over $\Sigma$ is a finite collection of documents over $\Sigma$ where each document has a name. Formally, a \emph{document database} is a function $D: \{d_1, \ldots, d_m\} \rightarrow \Sigma^*$ where $\{d_1, \ldots, d_m\}$ is a finite set of \emph{document names} and $D(d_i)$ is the document assigned to the name $d_i$. We also say that $\{d_1, \ldots, d_m\}$ is the \emph{schema} of the document database $D$. We define the size of $D$ as $|D| = \sum_{i=1}^m |D(d_i)|$.

Given a document database $D$ we can create new documents by a sequence of text-editing operations.
Here, we introduce the notion of a \emph{Complex Document Editing expression} (CDE-expression for short) over document names $\{d_1, \ldots, d_m\}$, which is defined by the following \emph{syntax}: 
\[ \renewcommand{\arraystretch}{1.2}
\begin{array}{rcl}
	\varphi & := &  d_\ell, \ell\in[1, m] \\
	& \mid &  \concat(\varphi, \varphi) \\
	& \mid &  \extract(\varphi, i, j) \\
	& \mid &  \delete(\varphi, i, j) \\
	& \mid &  \insertop(\varphi, \varphi, k) \\
	& \mid & \copyop(\varphi, i, j, k)
\end{array}
\]
where the values $i,j$ are valid {\it positions}, and $k$ is a valid {\it gap}. % -- concepts that we will explain shortly. 
We define $|\varphi|$ as the number of document names and operations in $\varphi$.

The \emph{semantics} of an CDE-expression $\varphi$ defines a function $\sem{\varphi}$ from document databases over $\{d_1, \ldots, d_m\}$ to documents, and works recursively as follows:
\[
\renewcommand{\arraystretch}{1.4}
\begin{array}{rcl}
	\sem{d_\ell}(D) & = & D(d_\ell) \\
	\sem{\concat(\varphi_1, \varphi_2)}(D) & = & \sem{\varphi_1}(D) \cdot \sem{\varphi_2}(D)  \\
	\sem{\extract(\varphi, i, j)}(D) & = & \sem{\varphi}(D)\spanc{i}{j+1} \\
	\sem{\delete(\varphi, i, j)}(D) & = & \sem{\varphi}(D)\spanc{1}{i} \cdot \sem{\varphi}(D)\spanc{j+1}{|\sem{\varphi}(D)|+1} \\
	\sem{\insertop(\varphi_1, \varphi_2, k)}(D) &=& \sem{\varphi_1}(D)\spanc{1}{k}\cdot \sem{\varphi_2}(D) \cdot \sem{\varphi_1}(D)\spanc{k}{|d|+1} \\
	\sem{\copyop(\varphi, i, j, k)}(D) &=& \sem{\varphi}(D)\spanc{1}{k}\cdot \sem{\varphi}(D)\spanc{i}{j+1} \cdot \sem{\varphi}(D)\spanc{k}{|\sem{\varphi}(D)|+1} 
\end{array}
\]
where $\varphi$, $\varphi_1$, and $\varphi_2$ are CDE-expressions, $i,j$ are valid positions, and $k$ is a valid gap. We assume that if $i$, $j$, or $k$ are not valid, then $\sem{\varphi}(D)$ is undefined. Verifying that positions are valid for $\varphi$ takes linear time in $|D|$ and $|\varphi|$ so, for the sake of simplification, we assume that positions and gaps are always valid. 

The complex document editing operations were formally introduced in~\cite{SchmidS22}, and here we present an equivalent formalization of the framework that better fits our purposes. As the name of each operation suggests: $\concat$ defines the concatenation of two documents, $\extract$ outputs the substring at the span $\spanc{i}{j+1}$ of a document, $\delete$ removes the substring at the span $\spanc{i}{j+1}$, $\insertop$ adds a document inside the other at position $k$, and $\copyop$ inserts the substring at the span $\spanc{i}{j}$ at the position $k$.
We refer the reader to~\cite{SchmidS22} for several examples and applications of CDE-expressions.

\paragraph{Multiroot SLPs}
A \emph{multiroot SLP} is a tuple $\rS = (N, \Sigma, R, \rS_0)$ where $N$ is a set of non-terminals, $\Sigma$ is the set of terminals, $R$ is a set of rules, and $\rS_0: \{d_1, \ldots, d_m\} \rightarrow N$ maps document names to non-terminals. We extend this notation so that $\rS(d_i) = (N, \Sigma, R, \rS_0(d_i))$ defines an SLP with start symbol $\rS_0(d_i)$. In other words, $\rS$ is like an SLP that additionally has multiple roots $\rS_0(d_1), \ldots, \rS_0(d_m)$. 

A multiroot SLP $\rS$ naturally defines a document database that maps each document name $d_i$ to $\doc(\rS(d_i))$. By some abuse of notation, we write $\doc(\rS)$ to denote the document database $\doc(\rS): \{d_1, \ldots, d_m\} \rightarrow \Sigma^*$ such that $[\doc(\rS)](d_i) = \doc(\rS(d_i))$.   Therefore, multiroot SLPs work as our formalism to specify compressed document databases where we can apply CDE-expressions. We naturally extend the semantics of a CDE-expression $\varphi$ from database documents to multiroot SLPs as $\sem{\varphi}(\rS) = \sem{\varphi}(\doc(\rS))$, assuming $\varphi$ is over the same schema $\{d_1, \ldots, d_m\}$.

\paragraph{The evaluation problem} 
We are ready to define the query evaluation problem of complex editing expressions over compressed document databases. Let $\mathcal{C}$ be any class of \crt and let $\mathcal{F}$ be a class of CDE-expressions (e.g., expressions that use a subset of the operations).
\vspace{0.5em}
\begin{center}
	\framebox{
		\begin{tabular}{rl}
			\textbf{Problem:} \!\!\!\!\!\!& $\enumedit[\mathcal{C}, \mathcal{F}]$  \\ \hline \vspace{-3mm} \\
			\textbf{Preprocessing:}  \!\!\!\!\!\! & 
			$\left\{ \text{\begin{tabular}{rl}
					\textbf{input:} & \!\!\! An \crt $\cA \in \mathcal{C}$ and a multiroot-SLP $\rS$ \\
					%                                \textbf{output:}
					\textbf{result:}
					& \!\!\! A data structure $\DS_{\cA,\rS}$
			\end{tabular}} \right.
			$            \\ \hline \vspace{-3mm} \\
			\textbf{Editing:}  \!\!\!\!\!\! & 
			$\left\{ \text{\begin{tabular}{rl}
					\textbf{input:} & \!\!\! The data structure $\DS_{\cA,\rS}$ and an CDE-expression $\varphi \in \mathcal{F}$ \\
					%                                \textbf{output:}
					\textbf{result:}
					& \!\!\! A data structure $\DS_{\cA,\rS, \varphi}$
			\end{tabular}} \right.
			$            \\ \hline \vspace{-3mm} \\
			\textbf{Enumeration:} \!\!\!\!\!\! &
			$\left\{
			\!\!\text{
				\begin{tabular}{rl}
					\textbf{input:} & \!\!\!
					The data structure $\DS_{\cA,\rS, \varphi}$ \\
					\textbf{output:} & \!\!\! Enumerate $\sem{\cA}(\sem{\varphi}(\doc(\rS)))$
				\end{tabular} 
			}\right.$\!\!\!\!
		\end{tabular}
	}
\end{center}

\vspace{0.5em}
The evaluation problem $\enumedit$ is an extension of the problem $\enumraa$ studied in the previous sections, in the sense that it also has preprocessing and enumeration phases that work as before. The main difference is that it has a middle phase, called \emph{editing}, where we can receive any CDE-expression $\varphi$ for evaluating the \crt $\cA$ over the new document $\sem{\varphi}(\doc(\cA))$. Specifically, the editing phase receives the data structure $\DS_{\cA, \rS}$ of the preprocessing phase plus a CDE-expression $\varphi$. Then, it produces a new data structure $\DS_{\cA, \rS, \varphi}$ that will be used later in the enumeration phase of $\sem{\cA}(\sem{\varphi}(\doc(\rS)))$ (i.e., it encodes the outputs). 
In other words, the task is to evaluate $\cA$ efficiently over any edited document without re-processing the data. 
We say that an algorithm for $\enumedit$ has \emph{$g$-editing time} if there exists a constant $c$ such that, for every input $(\cA, \rS, \varphi)$, the time for the editing phase of $\varphi$ is bounded by $c \cdot g(\cA, \rS, \varphi)$. Note that, for a fine-grained analysis of the editing phase's running time, we consider all the inputs: $\cA$, $\rS$, and $\varphi$. This is because the data structure $\DS_{\cA,\rS}$ depends on $\cA$ and $\rS$.

It is important to note that this setting is equivalent to the setting presented in~\cite{SchmidS22}. The main difference is in the presentation, where we are more explicit in the semantics of CDE-expressions and in formalizing the preprocessing, editing, and enumeration phases. In~\cite{SchmidS22}, the authors also considered that the input for the preprocessing phase can have several queries $\cA_1, \ldots, \cA_k$ that the user might want to evaluate simultaneously. For the sake of presentation, we adopted the setting with a single query (i.e., an \crt) since one can naturally extend our setting to $k$ queries with the cost of having an additional $k$ factor in the running time of the preprocessing and editing phases. 
One other subtlety that~\cite{SchmidS22} has is that their setting allows the user to edit $\DS_{\cA, \rS}$ with a CDE-expression $\varphi$ and then select any $d$ in the schema to enumerate the set $\sem{\cA}([\doc(\rS)](d))$. This use case is still considered by our setting by letting the user define $\varphi' = d$ as a second edit, and enumerate $\sem{\cA}(\sem{\varphi'}(\doc(\rS))) = \sem{\cA}([\doc(\rS)](d))$ with virtually no extra cost.
Finally, our setting has the advantage that we consider \crt as input, which is more general than in~\cite{SchmidS22}, where they were restricted to extended variable-set automata.

\paragraph{Efficient evaluation with complex editing} We are almost ready to present the main result of this section---we just need a few more definitions for multiroot SLPs: We say that a multiroot SLP $\rS = (N, \Sigma, R, \rS_0)$ is in {\it Chomsky normal form} (or just normal form) if every rule in $R$ is of form $X\to a$ or $X\to YZ$, where $a\in \Sigma$ and $X, Y, Z\in N$. Also, we say that $\rS$ is {\it strongly balanced} if it is in normal form and, for each rule $X\to YZ$, the value $\ord(Y) - \ord(Z)$ is either -1, 0 or 1, where $\ord(X)$ is the maximum distance from $X$ to any terminal in the derivation tree.  We define the {\it maximum intermediate document size} $|\max_{\varphi}(D)|$ induced by a CDE-expression $\varphi$ on a document database $D$ as the maximum size of $\sem{\psi}(D)$ for any sub-expression $\psi$ of $\varphi$ (i.e., any substring $\psi$ of $\varphi$ that matches the CDE syntax).
\begin{thm}\label{theo:edit-main-theorem}
	Let $\mathcal{C}$ be the class of all unambiguous \crts and let $\mathcal{F}$ be the class of all CDE-expressions. Assuming that the input multiroot SLP $\rS$ is strongly balanced, one can solve $\enumedit[\mathcal{C}, \mathcal{F}]$ with $|\cA|^3 \times |S|$-preprocessing time, $|\varphi| \times |\cA|^3  \times \log|\max_{\varphi}(\doc(\rS))|$-editing time, and output-linear delay.  
\end{thm}
The version of this result shown in~\cite{SchmidS22} was for eVA instead of \crt, and has logarithmic delay instead of output-linear delay. The proof combines the techniques of~\cite{SchmidS22} for editing SLPs and the machinery developed in our present work.
\begin{proof}
	Let $\rS = (N, \Sigma, R, \rS_0)$ be a strongly balanced multiroot SLP and $\cA = (Q, \Sigma, \oalph, \Delta, \qinit, F)$ be an unambiguous \crt. In a nutshell, the algorithm strategy is as follows: During the preprocessing phase, we construct the succinct \dsabbr $\cD$ and the set of matrices $\{M_X\}_{X \in N}$ associated with $\rS$ and $\cA$ (i.e., by running Algorithm~\ref{alg:evaluation}). Then, during the editing phase, we extend $\rS$ to a multiroot SLP $\rS'$ by adding new non-terminals $N'$ (and rules over $N'$) that represent partial documents of evaluating $\varphi$ with a special non-terminal $S_0'$ representing the document $\sem{\varphi}(\rS)$. For each new non-terminal $X' \in N'$, we construct its matrix $M_{X'}$ extending $\cD$ to $\cD'$. Finally, we enumerate the results with output-linear delay using $\cD'$ and the matrix $M_{S_0'}$ similarly to Algorithm~\ref{alg:evaluation}. 
	
	We start by showing how to do the extension of $\rS$ to $\rS'$ by using $\varphi$.
	In the following, we assume that for each nonterminal $X$ of $\rS$ for which its rule has the form $X\to YZ$, the values $|\doc(X)|$, $\ord(X)$, and nonterminals $Y$ and $Z$ are accessible in constant time. 
	All these values can be precomputed with a linear-time pass over $\rS$ during the preprocessing. We call $\rS$ along with constant-time access to these values {\it the basic data structure for} $\rS$.
	Furthermore, a multiroot SLP $\rS' = (N \cup N', \Sigma, R', \rS'_0)$ is called an \emph{extension} of $\rS$ if $R'(A) = R(A)$ for every $A\in N$. In this context, we call $N'$ the {\it set of new non-terminals}.
	
	The following result from~\cite{SchmidS22} will be instrumental in the proof.
	\begin{thmC}[{\cite[Theorem 4.3]{SchmidS22}}]\label{theo:cedits}
		Let $\rS$ be a strongly balanced multiroot SLP with document names $d_1, \ldots, d_m$. When given the basic data structure for $\rS$ and a CDE-expression $\varphi$ over $\{d_1, \ldots, d_m\}$, one can construct a strongly balanced extension $\rS'$ of $\rS$, along with its basic data structure, and a non-terminal $S_0'$ of $\rS'$ such that $\doc(S_0') = \sem{\varphi}(\rS)$.
		This construction takes time $\cO\big(|\varphi|\cdot \log|\max_{\varphi}(\doc(\rS))|\big)$.  In particular, the number of new non-terminals $|N'|$ is in $\cO\big(|\varphi|\cdot \log|\max_{\varphi}(\doc(\rS))|\big)$.
	\end{thmC}
	Now, let $\cD$ be the succinct \dsabbr and $\{M_X\}_{X \in N}$ the set of matrices associated to $\rS$ and $\cA$ after the preprocessing phase. Further, let $\tilde{R} = R' \setminus R$ be the set of new rules in $\rS'$ after applying Theorem~\ref{theo:cedits}. Consider the DAG that is induced by this set $\tilde{R}$. We can go over these new rules $X'\to Y'Z'$ in a bottom-up fashion, starting from the ones where $Y',Z'\in N$, and define the nodes $M_{X'}[p,q] = (M_{Y'}\otimes \shiftop(M_{Z'},|\doc(Y')|))[p,q]$ for each $p,q\in Q$. 
%	Then, we build the nodes $v_{X'}$ for each new non-terminal $X'$ as defined in lines 4-6 in Algorithm~\ref{alg:evaluation}. This will give us that $\sem{\cD'}(v_{X'}) = \sem{\cA}(\doc(X'))$. 
	The new nodes are created by application of the \dsabbr operations $\add$, $\union$, $\prod$ and $\shiftop$ which create the succinct \dsabbr $\cD'$ that defines the query data structure for $\rS'$ and $\cA$. The procedure takes $\cO(|Q|^3\cdot|N'|)$ time. By Theorem~\ref{theo:cedits}, this means that the editing time is in $\cO\big(|Q|^3 \cdot |\varphi|\cdot \log|\max_{\varphi}(\doc(\rS))|\big)$.
	Finally, we can use the matrix $M_{S_0'}$ and the nodes at entries $(p, q) \in I \times F$ to retrieve the outputs with output-linear delay during the enumeration phase.
\end{proof}
It is important to notice that we assume that $\rS$ is strongly balanced and use Theorem~4.3 from~\cite{SchmidS22} to bound the number of new nonterminals that are necessary to implement the operations $\extract$, $\delete$, $\insertop$, and $\copyop$ of CDE-expressions. In \cite{SchmidS22}, the strongly balanced condition was also needed for enforcing the log-delay during enumeration, which is not the case here. In particular, we can get the following result if we assume that our CDE-expressions only use the $\extract$ operation. 
\begin{cor}\label{cor:concat}
	Let $\mathcal{C}$ be the class of all unambiguous \crts and $\mathcal{F}$ the class of all CDE-expressions that only uses document names and $\concat$ operations. One can solve $\enumedit[\mathcal{C}, \mathcal{F}]$ with $|\cA|^3 \times |S|$-preprocessing time, $|\varphi| \times |\cA|^3 $-editing time, and output-linear delay.  
\end{cor}
\begin{proof}
	The proof follows in the same line as Theorem~\ref{theo:edit-main-theorem}. The difference is that we do not need Theorem~\ref{theo:cedits} since the concatenation operation can be done directly by adding a new non-terminal for each $\concat$ operation. In particular, the number of new non-terminals $|N'|$ is~in~$\cO(|\varphi|)$.
\end{proof}
	
	\section{Future work}\label{sec:conclusions}
	
	%!TEX root = ../main.tex

%We showed an enumeration algorithm which efficiently evaluates queries defined by annotated automata over SLP-compressed documents. This result implies algorithm for spanner evaluation on highly-compressed documents in which the document is compressed first, and then evaluated over the query.

%We believe that the algorithm that was presented has a simple presentation, and would not be majorly complicated to implement in practice. Other algorithms for regular spanners have been implemented, achieving good performance, so having a practical algorithm that receives a compressed document, or compresses the document itself, might allow much faster running times for the same problem.

One natural direction for future work is to explore which other compression schemes allow output-linear delay enumeration for evaluating annotated automata. To the best of our knowledge, the only model of compressed data in which spanner evaluation has been studied is SLPs. However, other models (such as some based on run-length encoding) offer better compression rates and may yield more desirable results in practice.
The run-length extension of SLPs was introduced in~\cite{RLSLP} and they are defined similarly, but with the added feature of allowing productions of {\em runs}, i.e., of the form $A\to B^k$. Exploring evaluation for AnnA in this model is a clear next step for our work.

Recently, Lohrey and Schmid~\cite{LohreyS24} extended the results from (the preliminary version of) this paper from SLP-compressed words to SLP-compressed trees with similar data-complexity time bounds. 
A natural question that arises here is whether the complex editing results from~\cite{SchmidS22} can also be applied to this scenario. 
The framework developed in~\cite{Niewerth18, Sarah22} seems to be especially well-suited for this, though it would need to be extended beyond single-node updates to handle complex editing.

We believe the \dsabbr data structure is interesting in its own right and worth investigating further to see how much it could be extended while still allowing output-linear delay enumeration. 
Another aspect we could check is whether there are enumeration results in other areas that could be improved using \dsabbr.

%Another aspect that might be improvable in Shift-ECS, or even the original ECS data structure, is the fact that output-linear delay allows an enumeration scheme where the user waits $O(|w|)$ time to see a single symbol of $w$. It would be interesting to find a more fine-grained enumeration bound which produces each output symbol in an output string with constant-delay.

Lastly, it would be interesting to study whether one could apply fast matrix multiplication techniques to Algorithm~\ref{alg:evaluation} to improve the running time to sub-cubic in the number of~states.

%One more direction is seeing if the extending the concisely annotated automata model complex models of computation, such as Tree Automata or Pushdown Automata allows for further improvements in known results related to spanner evaluation.  

	\bibliographystyle{alphaurl}
	\bibliography{extras/biblio}

\end{document}